\documentclass[12pt,a4paper]{iopart}
\usepackage[utf8]{inputenc}
\usepackage[english]{babel}
\usepackage{graphicx}
\usepackage{color}
\usepackage[unicode=true,pdfusetitle,bookmarks=true,bookmarksnumbered=false,bookmarksopen=false,breaklinks=false,pdfborder={0 0 0},backref=false,colorlinks=true]{hyperref}
\usepackage{graphicx}
\usepackage{url}
\usepackage{amssymb,amsthm,amsfonts,amstext}
\expandafter\let\csname equation*\endcsname\relax
\expandafter\let\csname endequation*\endcsname\relax
\usepackage{mathtools} 

\usepackage{tikz}
\usepackage{tikz-3dplot}
\usetikzlibrary{calc}

\usepackage{subcaption}

\newcommand{\corr}{\mathrm{corr}}
\newcommand\ii{{\mathrm{i}}}

\newcommand{\expect}[1]{\langle #1\rangle}
\tikzset{
world/.style={circle,draw,minimum size=0.55cm,fill=gray!15},
world2/.style={circle,draw,minimum size=1.2cm,fill=gray!15},
}
\newcommand{\fidelity}{\text{F}}
\DeclareMathOperator{\dist}{dist}
\newcommand{\purity}{\mathcal{P}}

\begin{document}

\title[Locality of temperature and correlations for non-zero-temperature PT]{Locality of temperature and correlations in the presence of non-zero-temperature phase transitions}
\author{Senaida Hern\'andez Santana}
\address{Technical University of Madrid, 28040 Madrid, Spain}
\address{ICFO-Institut de Ciencies Fotoniques, The Barcelona Institute of Science and Technology, 08860 Castelldefels (Barcelona), Spain}
\author{Andr\'as Moln\'ar}
\address{Complutense University of Madrid, 28040 Madrid, Spain}
\author{Christian Gogolin}
\address{Covestro  Deutschland  AG,  Kaiser  Wilhelm  Allee  60,  51373  Leverkusen,  Germany}
\address{ICFO-Institut de Ciencies Fotoniques, The Barcelona Institute of Science and Technology, 08860 Castelldefels (Barcelona), Spain}
\author{J. Ignacio Cirac}
\address{Max-Planck-Institut für Quantenoptik, Hans-Kopfermann-Straße 1, 85748 Garching, Germany}
\author{Antonio Ac\'in}
\address{ICFO-Institut de Ciencies Fotoniques, The Barcelona Institute of Science and Technology, 08860 Castelldefels (Barcelona), Spain}
\address{ICREA-Institucio Catalana de Recerca i Estudis Avan\c{c}ats, Pg. Lluis Companys 23, 08010 Barcelona, Spain}

\begin{abstract}
While temperature is well understood as an intensive quantity in standard thermodynamics, it is less clear whether the same holds in the presence of strong correlations, especially in the case of quantum systems, which may even display correlations with no classical analogue.
The problem lies in the fact that, under the presence of strong correlations, subsystems of a system in thermal equilibrium are, in general, not described by a thermal state at the same temperature as the global system and thus one cannot simply assign a local temperature to them.
However, there have been identified situations in which correlations in thermal states decay sufficiently fast so that the state of their subsystems can be very well approximated by the reduced states of equilibrium systems that are only slightly bigger than the subsystems themselves, hence allowing for a valid local definition of temperature.
In this work, we address the question of whether temperature is locally well defined for a bosonic system with local interactions that undergoes a phase transition at non-zero temperature.
We consider a three-dimensional bosonic model in the grand canonical state and verify that a certain form of locality of temperature holds regardless of the temperature, and despite the presence of infinite-range correlations at and below the critical temperature of the phase transition.

\end{abstract}

\maketitle

\section{Introduction}

Understanding how thermodynamic processes are affected by the presence of quantum phenomena is a topic of intense research since the early days of quantum physics~\cite{Gemmer}.
For instance, it is still not fully understood how thermodynamic quantities, many of them originally presented as statistical magnitudes, should be defined in microscopic systems where fluctuations become important or whether their properties differ from those in classical systems. More in general, the goal is to understand if and how thermodynamic concepts can be extended beyond the, often idealized, conditions in which they were originally derived.
One such example is temperature and its intensive character in standard thermodynamics: given a system in equilibrium with a well-defined temperature, one can assign an equilibrium state at the same temperature to any part of it.
However, this property valid in standard thermodynamics does not generally hold in the presence of strong correlations. This is the question we consider in this work, in particular in the context of quantum systems, which may display forms of correlations with no classical analogue.
We study under what conditions temperature can be locally well defined for subsystems of a global system at a well-defined temperature.

A standard setting to study this problem is given by a system at thermal equilibrium at temperature $T$ and described by a local Hamiltonian $H$ that is a sum of local interacting terms.
Specifically, the system is at a thermal state at inverse temperature $\beta:=1/k_B T$, with the expression
\begin{equation}\label{eq:defthermalstate}
  \Omega_{\beta}[H] := \frac{\e^{-\,\beta\,H}}{Z_{\beta}[H]},
\end{equation}
where $Z_{\beta}[H]$ is the canonical partition function  $Z_{\beta}[H] :=  \tr(\e^{-\,\beta\,H})$.
The main question we address here is to understand whether it is possible to assign to its subsystems a thermal description at the same temperature $T$.
If no constraints are imposed on the Hamiltonian of the subsystem, this is always possible: it suffices to define $H=-\log(\rho)/\beta$, where $\rho$ is the state of the subsystem. Nevertheless, this solution is not satisfactory, as the resulting Hamiltonian is in general arbitrary and has no physical interpretation.
In the case of systems with local interactions, however, there is a natural choice for the subsystem Hamiltonian defined by the local terms with support on the subsystem.
For this choice of local Hamiltonian, the reduced state of a thermal state cannot be typically expressed by Eq.~\ref{eq:defthermalstate} (not even approximately).

In some cases, the global thermal state can be very well approximated by the product of local thermal states and one can use a local definition of temperature. This occurs, for instance, for some models with weak interactions. However, that is typically not the case for strong interactions, where in general it is not straightforward to define a local thermal state and corresponding temperature, since this depends on the environment, that is, the rest of the system, and the interactions that couple the subsystem to it.

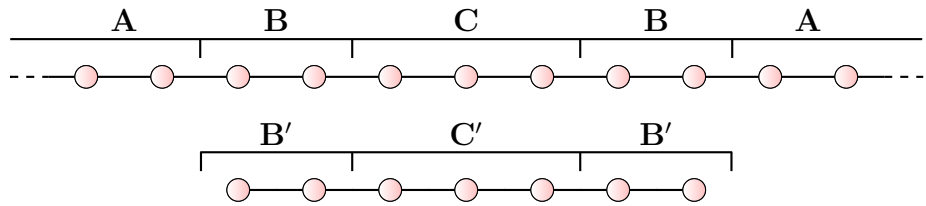
\begin{figure}[t!]
  \centering
\begin{tikzpicture}
\draw [-,thick] (-6,0.5) -- (6,0.5);
\draw [-,thick] (-3.5,0.5) -- (-3.5,0.25);
\draw [-,thick] (-1.5,0.5) -- (-1.5,0.25);
\draw [-,thick] (1.5,0.5) -- (1.5,0.25);
\draw [-,thick] (3.5,0.5) -- (3.5,0.25);
    \draw[style=help lines, black,thick] (-5.5,0) grid[step=1cm] (5.5,0);
    \draw[style=help lines, black,thick,dashed] (5.5,0) grid[step=1cm] (6,0);
    \draw[style=help lines, black,thick,dashed] (-6,0) grid[step=1cm] (-5.5,0);

    \foreach \x in {-5,...,5}{
      \foreach \y in {0,0}{
        \node[draw,circle,inner sep=3pt,fill, shade, left color=white,right color=pink] at (\x,\y) {};
      }
    }
\node at (0,1-0.25) {$\mathbf{C}$};
\node at (2.5,1-0.25) {$\mathbf{B}$};
\node at (-2.5,1-0.25) {$\mathbf{B}$};
\node at (4.5,1-0.25) {$\mathbf{A}$};
\node at (-4.5,1-0.25) {$\mathbf{A}$};

\draw [-,thick] (-3.5,-1.5+0.5) -- (3.5,-1.5+0.5);
\draw [-,thick] (-3.49,-1.5+0.5) -- (-3.49,-1.75+0.5);
\draw [-,thick] (-1.5,-1.5+0.5) -- (-1.5,-1.75+0.5);
\draw [-,thick] (1.5,-1.5+0.5) -- (1.5,-1.75+0.5);
\draw [-,thick] (3.49,-1.5+0.5) -- (3.49,-1.75+0.5);
    \draw[style=help lines, black,thick] (-3,-1.5) grid[step=0.5cm] (3,-1.5);
    \foreach \x in {-3,...,3}{
      \foreach \y in {-2+0.5,-2+0.5}{
        \node[draw,circle,inner sep=3pt,fill, shade, left color=white,right color=pink] at (\x,\y) {};
      }
    }

\node at (0,-1-0.25+0.5) {$\mathbf{C}'$};
\node at (2.5,-1-0.25+0.5) {$\mathbf{B}'$};
\node at (-2.5,-1-0.25+0.5) {$\mathbf{B}'$};
\end{tikzpicture}
    \caption{\label{fig:introsetting} Scheme of the one-dimensional setting for the problem of locality of temperature.
The boundary size of $\mathbf{B}$ and $\mathbf{B}'$ must be large enough to obtain a good approximation of the partial state $\rho_\mathbf{C}$ by computing $\rho_{\mathbf{C}'}$.}
\end{figure}

A way to tackle this problem for systems with local interactions was presented by Ferraro et al. in~\cite{ferraro2012intensive}.
They studied a system of coupled harmonic oscillators on a lattice and proposed a way to assign a temperature to a subsystem $\mathbf{C}$ by dividing the global system into three regions: the subsystem $\mathbf{C}$, for which we would like to provide a thermal description, a boundary region $\mathbf{B}$ around it, and the rest of the system $\mathbf{A}$ (see Figure \ref{fig:introsetting}).
Given this setup, they considered a different system at thermal equilibrium at temperature $T$ which consists of two regions $\mathbf{C}'$ and $\mathbf{B}'$ that are of the same size and have the same type of interactions as the subsystem $\mathbf{C}$ and its boundary $\mathbf{B}$, respectively.
Then, they showed that it is sufficient to compute the partial state of the subsystem $\mathbf{C}'$ to obtain a very good approximation of the state of the subsystem $\mathbf{C}$, which we refer to as effective thermal state.
If the size of the boundary region $\mathbf{B}$ needed to attain a given error in the approximation is independent of the total system size, temperature is said to be local.

In last years, there have been several works studying the question of the existence of local temperatures using the same approach as Ferraro et al. \cite{ferraro2012intensive}, see also~\cite{artur}.
These works proved the validity of the locality of temperature for different systems:
(i) 1D and 2D systems made of interacting harmonic oscillators \cite{ferraro2012intensive};
(ii) fermionic and spin systems with arbitrary dimension and at high enough temperature \cite{Kliesch2014};
(iii) finite-spin one-dimensional systems at arbitrary temperatures \cite{1367-2630-17-8-085007}.
This last result in fact shows that a valid local thermal description is even possible at quantum phase transitions in one-dimensional systems provided that the boundary region is large enough. 
Now, recall that all phase transitions in one-dimensional systems take place at zero temperature.
However, this is no longer the case for higher dimensional systems, where phase transitions may also occur at non-zero temperature.
These phase transitions at non-zero temperature represent interesting phenomena in many-body systems and highly influence how correlations decay, a question which in turn is closely linked to the locality of temperature \cite{Kliesch2014}.
Our aim is to understand how phase transitions at non-zero temperature affect the possible local definition of temperature.

In this work, we tackle this problem for a three-dimensional bosonic system at thermal equilibrium at temperature $T$ and at particle density $n$.
This system undergoes a phase transition at a non-zero critical temperature $T_c$: a Bose-Einstein condensate is formed, i.e., a macroscopic fraction of particles have zero momentum below the critical temperature.
We observe that temperature is local above the critical temperature ($T>T_c$) and pseudo-local, as defined in what follows, below and at the critical temperature ($T\leq T_c$).
We also compare these results to the structure of spatial correlations of the system and find a good agreement between local temperature and correlations independently of the presence of the phase transition.

\section{Model}\label{sec:model}
We consider a free bosonic system in a three-dimensional lattice of volume $L^3$ with periodic boundary conditions (PBC) that is described by the Hamiltonian
\begin{equation}\label{eq:BEmodel}
  H := -\sum_{\langle\textbf{n},\,\textbf{m}\rangle} b^\dagger_\textbf{m}\,b_\textbf{n} + 6\sum_\textbf{n} b^\dagger_\textbf{n}\,b_\textbf{n},
\end{equation}
where $b$ and $b^\dagger$ are bosonic creation and annihilation operators, and $\textbf{n}=(n_x,n_y,n_z)$ with $n_i \in [-L/2,L/2-1]$, and $i\in\{x,y,z\}$.
This Hamiltonian describes the discretized version of the ideal Bose gas. It can be diagonalized by Fourier transform $\tilde b_\textbf{n}=\frac {1}{L^{3/2}} \sum_{\textbf{k}} b_\textbf{k}\,\e^{\ii\,\textbf{k}\,\textbf{n}}$, obtaining its  eigenvalues $\varepsilon(\textbf{k})$, equal to
\begin{equation}\label{eq:PBCeig}
  \varepsilon(\textbf{k}) := 2\,(3-\cos(k_x)-\cos(k_y)-\cos(k_z)).
\end{equation}
In three dimensions, this model has a phase transition at a non-zero temperature $T_c$ for fixed particle density $n := \expect{N}/V$, as below this temperature the particles form a Bose-Einstein condensate, i.e. a macroscopic fraction of the particles are in the zero momentum mode \cite{Pitaevskii2016}.
In order to fix both temperature and particle density, we consider a system in the grand canonical ensemble,
\begin{equation}\label{eq:GCstate}
  \Omega_{\{\beta,\,\mu\}} := \e^{-\,\beta\,H+\mu\,N} / \tr(\e^{-\,\beta\,H+\mu\,N})
\end{equation}
We choose the chemical potential $\mu=\mu(\beta)$ such that the particle density $n$ becomes fixed at $n=1$. We numerically estimate that the critical temperature is around $T_c \approx 5.6$ (see \ref{app:phase_transition}).

\section{Locality of temperature}\label{sec:locality}
In this section, we analyze whether the temperature can be locally well defined in the considered system and how this depends on the presence of a phase transition at non-zero temperature.
\subsection{The problem}

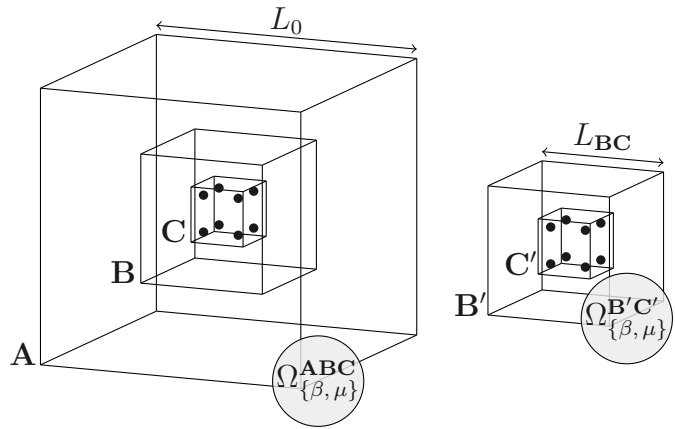
\begin{figure}
  \centering
  \newcommand{\drawcube}[2]{
    \coordinate (center) at #1 {};
    \foreach[count=\i] \phi in {45,135,-135,-45}
    {
      \coordinate (up\i) at ($(center)+(\phi:{#2/sqrt(2)})+(0,0,-#2*0.5)$) {};
      \coordinate (down\i) at ($(center)+(\phi:{#2/sqrt(2)})+(0,0, #2*0.5)$) {};
      \draw (up\i) -- (down\i);
    }
    \foreach[count=\i] \phi in {45,135,-135,-45}
    {
      \pgfmathsetmacro{\iplus}{mod(\i,4)+1}
      \draw (up\i) -- (up\iplus);
      \draw (down\i) -- (down\iplus);
    }
  }
  \tdplotsetmaincoords{78}{114}
  \begin{tikzpicture}[tdplot_main_coords,scale=0.5]

    \drawcube{(0,0,0)}{1.5}
    \drawcube{(0,0,0)}{3.5}
    \drawcube{(0,0,0)}{7.5}
    \drawcube{(0,10,0)}{1.5}
    \drawcube{(0,10,0)}{3.5}

    \foreach \x in {-0.5,...,0.5}
    \foreach \y in {-0.5,...,0.5}
    \foreach \z in {-0.5,...,0.5}
    \fill[black!90!white]  (\x,\y,\z) circle (0.3em);

    \foreach \x in {-0.5,...,0.5}
    \foreach \y in {9.5,...,10.5}
    \foreach \z in {-0.5,...,0.5}
    \fill[black!90!white]  (\x,\y,\z) circle (0.3em);

\pgfsetfillopacity{0.8}

\node[world2] at (3.75,3.75+0.5,-3.75+0.25) {};
\node[world2] at (1.75,10+1.75+0.5,-1.75+0.25) {};
\pgfsetfillopacity{0.9}

\node at (3.75,3.75+0.5,-3.75+0.25) {$\Omega_{\{\beta,\,\mu\}}^{\mathbf{ABC}}$};
\node at (1.75,10+1.75+0.5,-1.75+0.25) {$\Omega_{\{\beta,\,\mu\}}^{\mathbf{B'C'}}$};
\node at (3.75,-3.75-0.5,-3.75+0.25) {$\mathbf{A}$};
\node at (1.75,-1.75-0.5,-1.75+0.25) {$\mathbf{B}$};
\node at (0.75,-0.75-0.5,-0.75+0.25) {$\mathbf{C}$};
\node at (1.75,10-1.75-0.5,-1.75+0.25) {$\mathbf{B'}$};
\node at (0.75,10-0.75-0.5,-0.75+0.25) {$\mathbf{C'}$};
\node at (-3.75,0,3.75+0.6+0.1) {$L_0$};
\node at (-1.75,10,1.75+0.6+0.2) {$L_\mathbf{BC}$};


\path[<->](-3.75,-3.75,3.75+0.25) edge (-3.75,+3.75,3.75+0.25);
\path[<->](-1.75,10-1.75,1.75+0.25) edge (-1.75,10+1.75,1.75+0.25);

  \end{tikzpicture}
    \caption{\label{fig:setting} Left: Setting of the subsystem of interest $\mathbf{C}$, the boundary region $\mathbf{B}$ and their environment $\mathbf{A}$.
Right: Setting of the reference system with the subsystem $\mathbf{C}'$ and the boundary $\mathbf{B}'$, which are of the same size as $\mathbf{C}$ and $\mathbf{B}$ respectively.}
\end{figure}

We consider the state \eqref{eq:GCstate} on two cubic systems $\mathbf{ABC}$ and $\mathbf{B'C'}$ such that subsystems $\mathbf{C}$ and $\mathbf{C'}$ have size $L_C$, and $\mathbf{B}$ and $\mathbf{B'}$ size $L_{BC}$ (see Figure \ref{fig:setting}). The reduced density states of the regions $\mathbf{C}$ and $\mathbf{C'}$ are $\rho_C$ and $\rho_{C'}$, respectively.

If $\rho_C \approx \rho_{C'}$, observables acting on these subsystems (such as a thermometer) cannot distinguish between the global system $\mathbf{ABC}$ and the smaller system $\mathbf{B'C'}$. If this happens for a small boundary size $L_{BC}-L_C$, it means that temperature can be defined locally. We therefore compare the states $\rho_C$ and $\rho_{C'}$ while keeping the size of $\mathbf{ABC}$ fixed at a value $L_0$, the size of $\mathbf{C}$ and $\mathbf{C'}$ fixed at $L_C=2$, and varying the size $L_{BC}$. This comparison is made at temperatures both above and below the phase transition to understand how a non-zero-temperature phase transition affects the locality of  temperature.

\subsection{Methods}
In order to compare the states $\rho_\mathbf{C}$ and $\rho_{\mathbf{C}'}$,  we use the quantum fidelity \cite{mikeike}, a measure of distinguishability between any two quantum states $\rho$ and $\sigma$  defined as
\begin{equation*}
  \fidelity(\rho,\sigma) := \left[\tr\left(\sqrt{\rho^{1/2}\,\sigma\,\rho^{1/2}}\right)\right]^2 .
\end{equation*}
The fidelity satisfies $\fidelity(\rho,\sigma)\in[0,1]$, and $\fidelity(\rho,\sigma)=1$ if, and only if $\rho=\sigma$.

All the states of our interest, namely the Grand Canonical state of a quadratic Hamiltonian and reduced states of it, are Gaussian. The elements of the covariance matrix of the Grand Canonical state $\Omega_{\{  \beta ,\,\mu  \}}$ (\refeq{eq:GCstate}) described by the Hamiltonian (\refeq{eq:BEmodel}), with size $L^3$, can be expressed as
\begin{gather}
    \label{eq:covmatfin}
    \expect{b^\dagger_\textbf{n}\,b_\textbf{m}}_{\{\beta,\,\mu\}} = \frac{1}{\sqrt{L^3}} \sum_{\textbf{k}} \frac{\e^{-\,\ii\,(\textbf{n}-\textbf{m})\,\textbf{k}}}{\e^{\beta\,\varepsilon(\textbf{k})+\mu}-1}, \\
    \label{eq:covmat1}
  \expect{b_\textbf{n}\,b^\dagger_\textbf{m}}_{\{\beta,\,\mu\}} = \delta_{\textbf{m},\textbf{n}}+\expect{b^\dagger_\textbf{n}\,b_\textbf{m}}_{\{\beta,\,\mu\}},\\
\label{eq:covmat2}
  \expect{b_\textbf{n}\,b_\textbf{m}}_{\{\beta,\,\mu\}} = \expect{b^\dagger_\textbf{n}\,b^\dagger_\textbf{m}}_{\{\beta,\,\mu\}} = 0,
\end{gather}
where $\textbf{k}$ is the momentum vector and $\varepsilon(\textbf{n})$ is the eigenvalue of the Hamiltonian \eqref{eq:PBCeig}.
The covariance matrix of any of its subsystems is obtained by taking the matrix elements corresponding to the sites within that subsystem.
At last, we compute the fidelity between the two resulting Gaussian states, $\rho_\mathbf{C}$ and $\rho_{\mathbf{C'}}$, from their covariance matrices using the formula of Paraoanu et al.~\cite{Paraoanu1999a}.

\subsection{Results}

\begin{figure}
\centering
\includegraphics[width=0.5\textwidth]{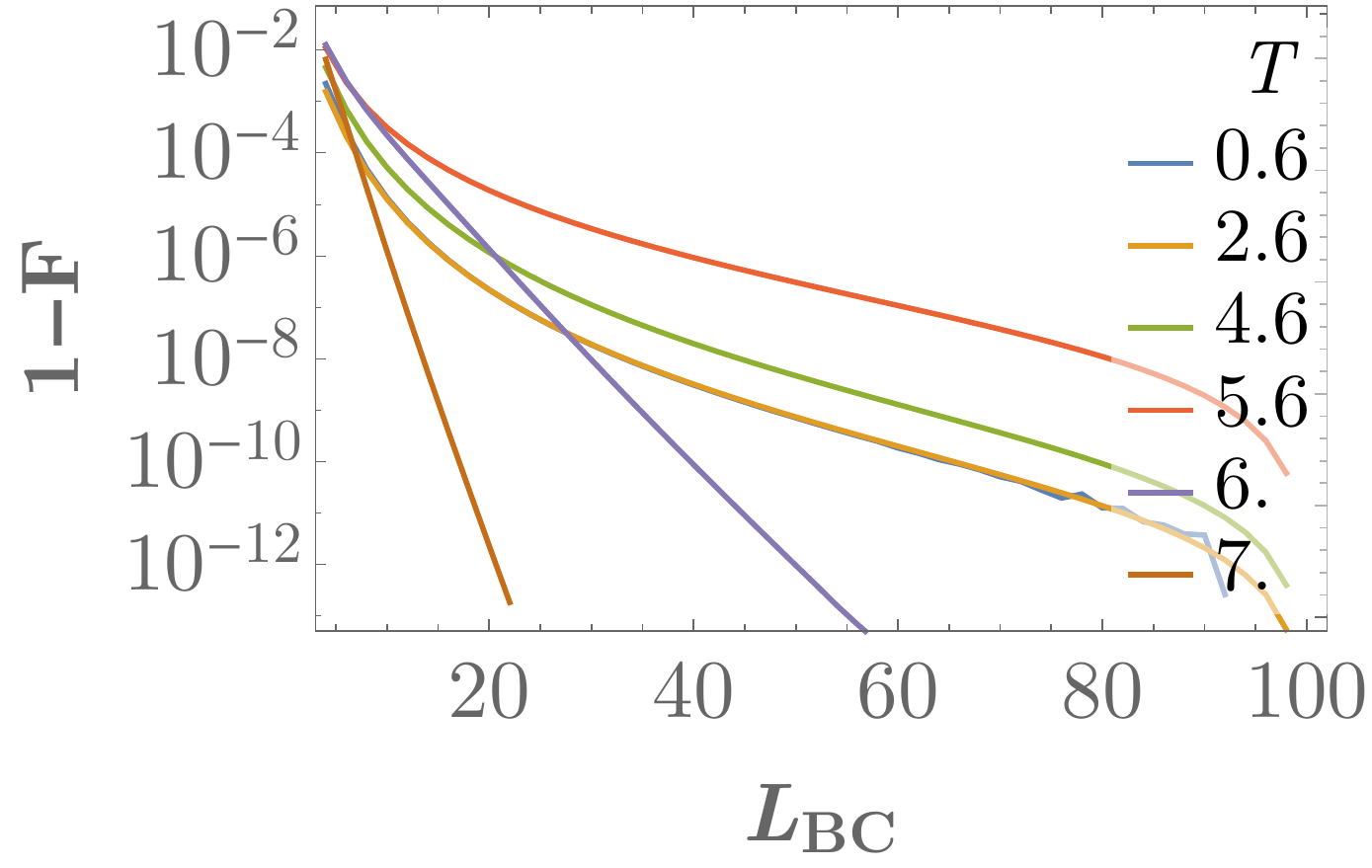}
\caption{\label{fig:fidelityvsboundary}Fidelity, $\fidelity(\rho_\mathbf{C},\rho_\mathbf{C'})$, vs reference system length, $L_\mathbf{BC}$, for different temperatures $T=0.6, 2.6, 4.6, 5.6, 6, 7$ and system size $L_0=100$.}
\end{figure}

We have plotted the obtained values of $1-F(\rho_C,\rho_{C'})$ as a function of $L_{BC}$ for different temperatures and for a system size $L_0=100$ (see Figure \ref{fig:fidelityvsboundary}). Calculations are in principle possible for larger system sizes as the relevant states are Gaussian, but we observe that the results show almost no variations for systems such as $L_0\geq 60$. We therefore expect that computations for $L_0=100$ provide a good approximation of the thermodynamic limit, with a reasonable numerical effort.
First, we observe that the fidelity between the states $\rho_\mathbf{C}$ and $\rho_\mathbf{C'}$ increases monotonically with $L_{BC}$, i.e. with the size of the boundary region $\mathbf{B}$.
We discuss below its different behaviours depending on whether we are below or at the critical temperature, $T\leq T_c$, or above, $T>T_c$.

\begin{figure}[t!]
\centering
\begin{subfigure}[t]{0.48\textwidth}
  \centering
  \includegraphics[width=0.97\textwidth]{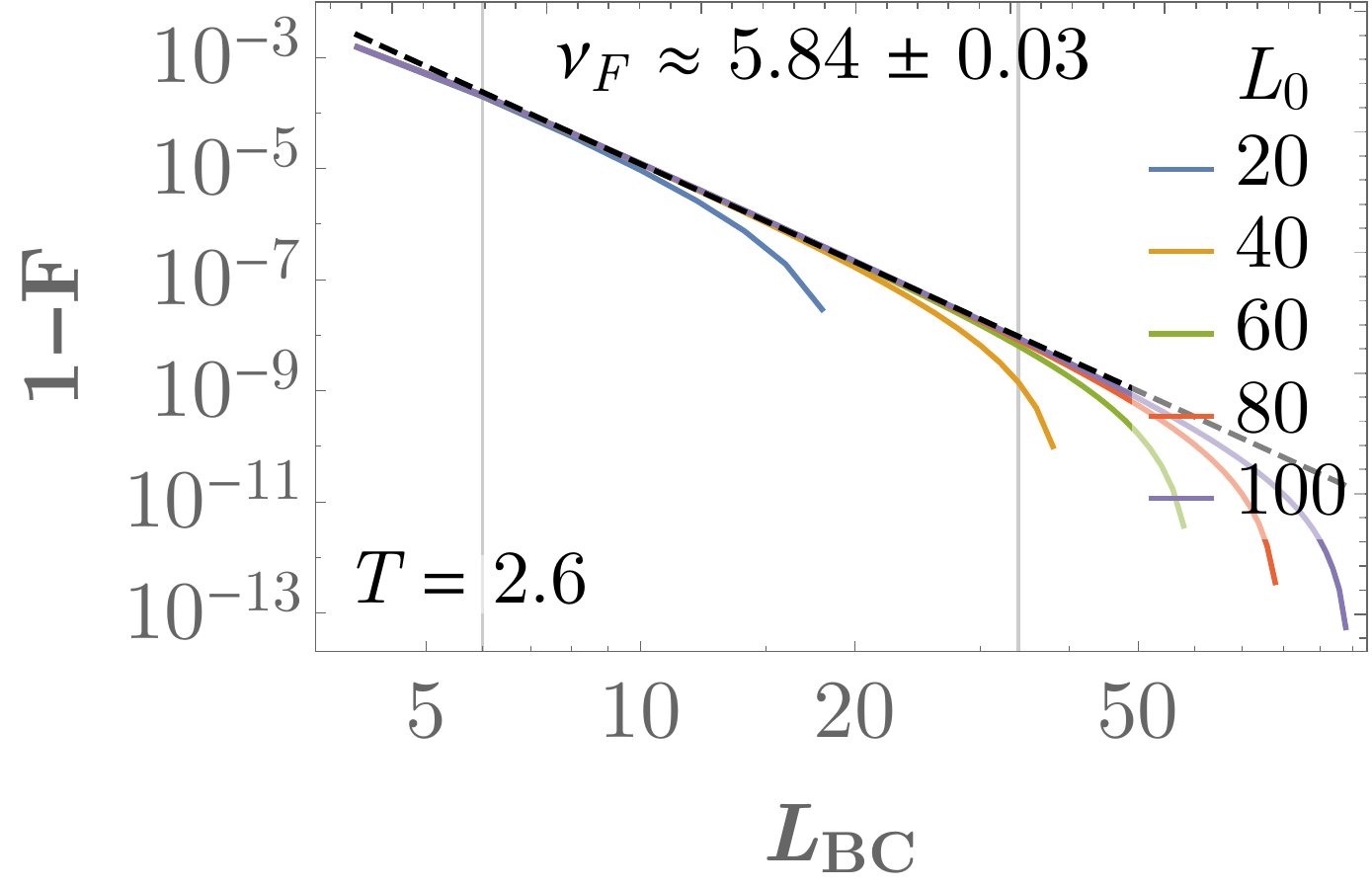}
  \caption{Fidelity error, $1-\fidelity(\rho_\mathbf{C},\rho_\mathbf{C'})$, vs length $L_\mathbf{BC}$ for different system lengths $L_0$ and temperature $T=2.6$, below the critical temperature $T_c\approx 5.6$. The dashed line corresponds to a power-law fit for $L_0=100$ and $L_\mathbf{BC}\in[6,L_0/3]$.}
\end{subfigure}\hfill
\begin{subfigure}[t]{0.48\textwidth}
  \centering
  \includegraphics[width=0.92\textwidth]{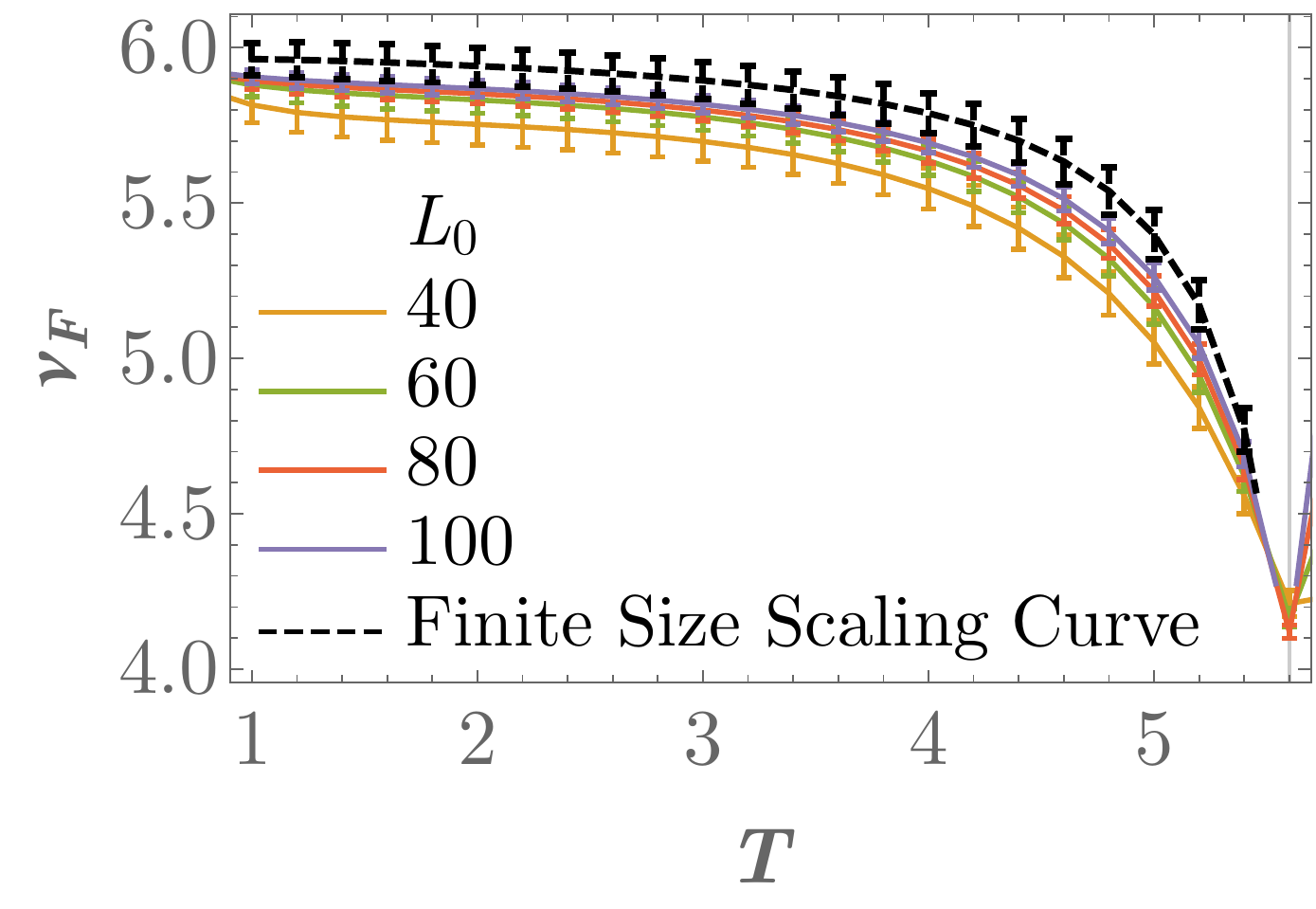}
  \caption{Exponent $\nu_F$ for power-law fit \eqref{eq:powerlawfit} vs temperature below the phase transition ($T\leq T_c$).
  The dashed line corresponds to the curve resulting of finite-size analysis.}
\end{subfigure}\\
\begin{subfigure}[t]{0.48\textwidth}
  \centering
  \includegraphics[width=0.97\textwidth]{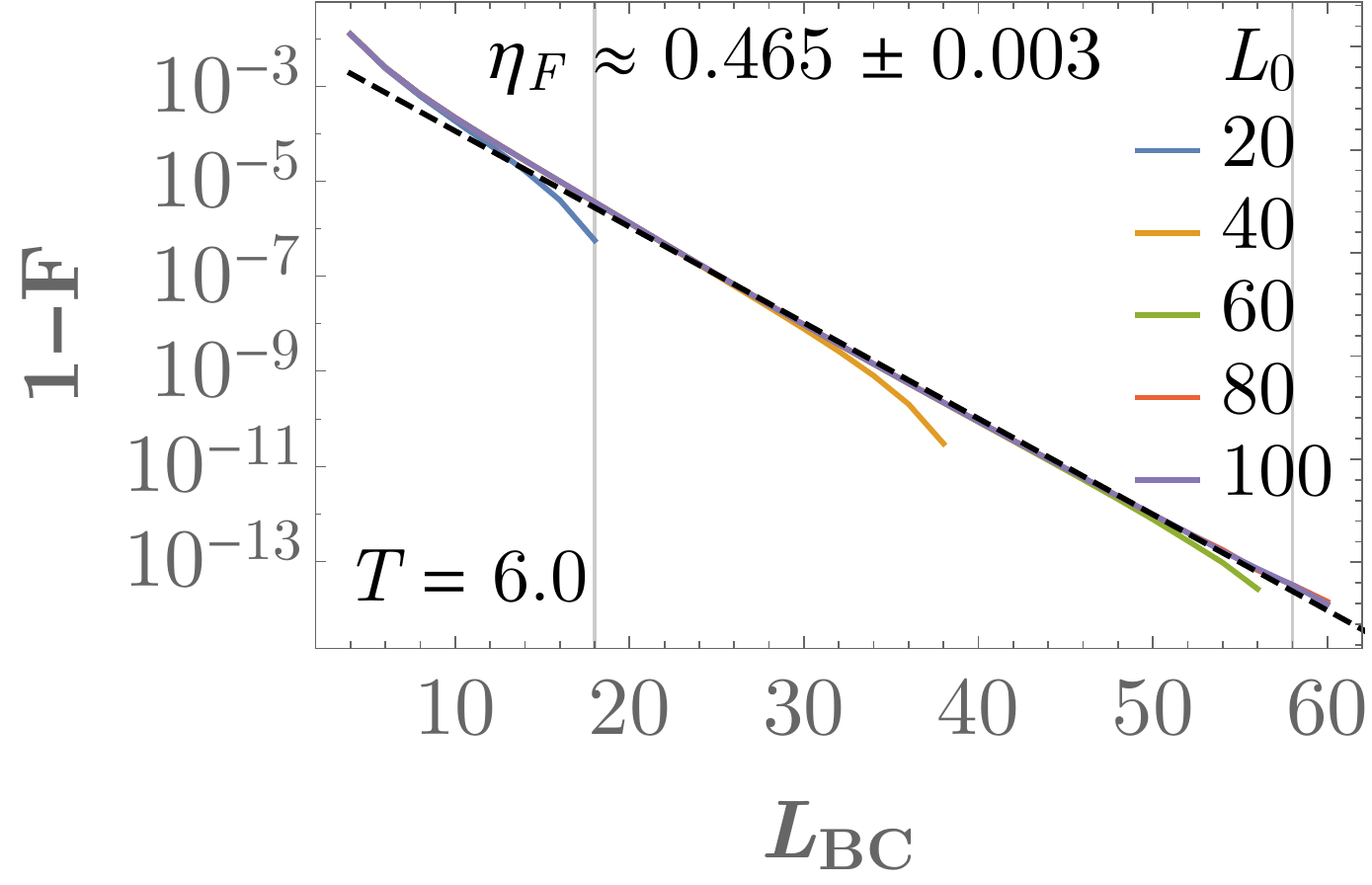}
  \caption{Fidelity error, $1-\fidelity(\rho_\mathbf{C},\rho_\mathbf{C'})$, vs length $L_\mathbf{BC}$ for different systems lengths $L_0$ and temperature $T=6$, above the critical temperature $T_c$.
  The dashed line corresponds to an exponential fit \eqref{eq:exponentialfit} for $L_0=100$ and data in the interval $[L_\mathbf{BC}^{\text{min}},L_\mathbf{BC}^{\text{max}}]$, where $L_\mathbf{BC}^{\text{max}}=\min (2\,L_0/3,L_{\text{cut}})$ with $L_{\text{cut}}=\{L \in \mathbb R : \text{F}(L)\approx 1-10^{-14}\}$, and $L_\mathbf{BC}^{\text{min}}=\max (L_\mathbf{BC}^{\text{max}}-2\,L_0/5,6)$.}
\end{subfigure}\hfill
\begin{subfigure}[t]{0.48\textwidth}
  \centering
  \includegraphics[width=0.92\textwidth]{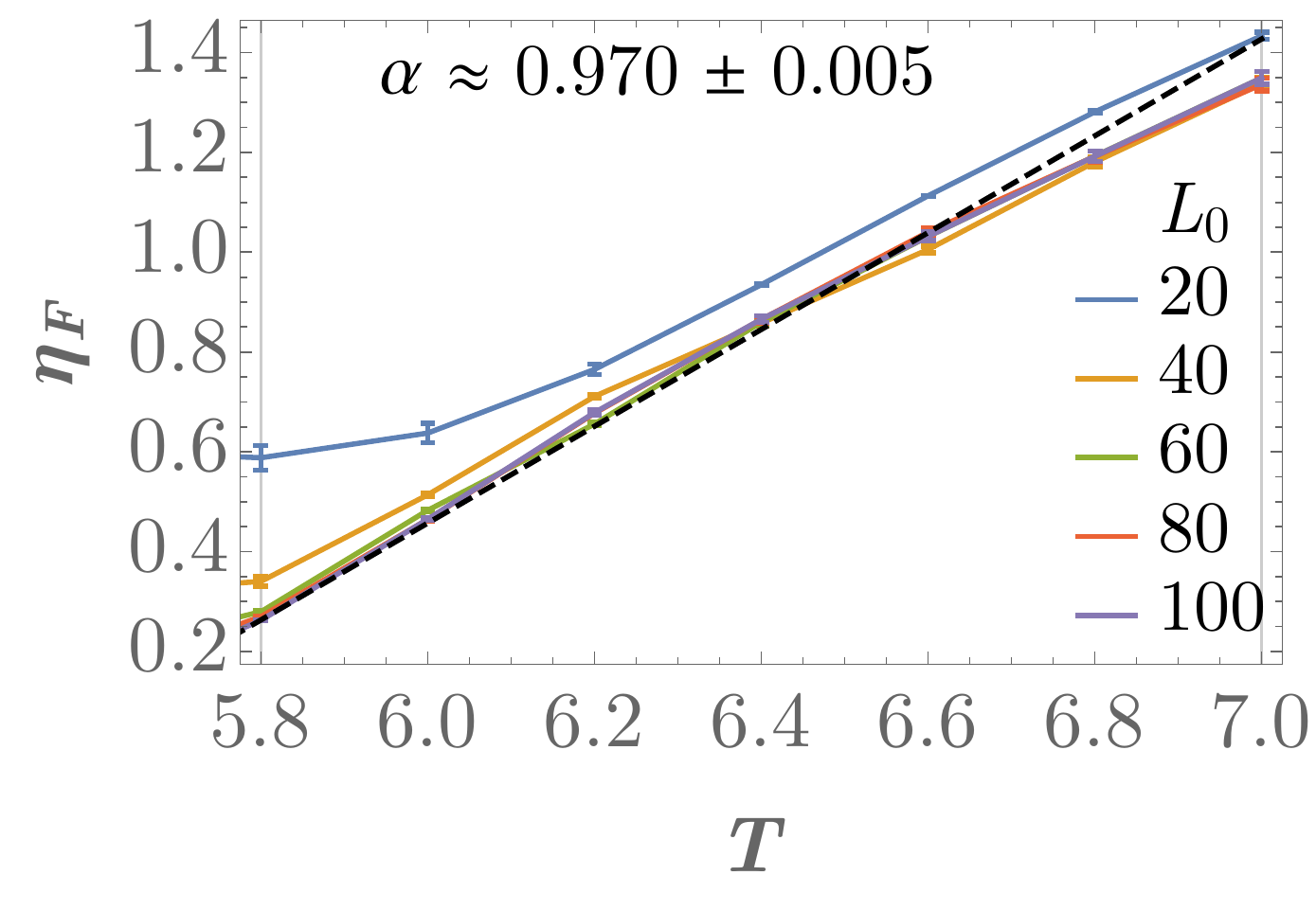}
  \caption{Exponent $\eta_F$ for exponential fit vs temperature above the phase transition ($T>T_c$).
  The dashed line corresponds to the linear fit of data for $L_0=100$ with exponent $\alpha$.
  Notice that the fidelity decreases polynomially for $T\leq T_c$ and exponentially for $T>T_c$.
  Error uncertainty and error bars represent the standard error of the parameters, as obtained from the linear interpolations.}
\end{subfigure}
\caption{\label{fig:fidelityvsboundary2} Fidelity vs system size and its scaling above and below phase transition}
\end{figure}

Below and at the phase transition the fidelity
, $\fidelity(\rho_\mathbf{C},\rho_\mathbf{C'})$, increases polynomially to $\fidelity=1$ with the length $L_\mathbf{BC}$ (see Figure \ref{fig:fidelityvsboundary2}$.\text{(a)}$), described by
\begin{equation}\label{eq:powerlawfit}
  1-\fidelity(\rho_\mathbf{C},\rho_\mathbf{C'}) \propto \frac{1}{L_\mathbf{BC}^{\nu_F}},
\end{equation}
with exponent $\nu_F$.
We analyze how the exponent, $\nu_F$, behaves as a function of the temperature, $T$, for different system sizes $L_0$ and make a finite-size analysis (see Figure~\ref{fig:fidelityvsboundary2}$.\text{(a)}$ and \ref{fig:fidelityvsboundary2}$.\text{(b)}$).
At the continuous limit, we observe that the exponent remains more or less constant around a value slightly below $6$ for temperatures $T\leq 4$, away from the phase transition, and that it suddenly decreases up to $\nu_F \approx 4$ around the phase transition, at $4<T<5.6$, see Figure~\ref{fig:fidelityvsboundary2}$.\text{(b)}$.
The exponents for finite systems were computed by fitting the data of the error fidelity, $1-\fidelity(\rho_\mathbf{C},\rho_\mathbf{C'})$, in the range $L_\mathbf{BC}\in[6,L_0/3]$.

Above the phase transition we find that the fidelity increases exponentially to $\fidelity=1$ with $L_\mathbf{BC}$ (see Figure \ref{fig:fidelityvsboundary2}$.\text{(c)}$), that is,
\begin{equation}\label{eq:exponentialfit}
  1-\fidelity(\rho_\mathbf{C},\rho_\mathbf{C'}) \propto \e^{-\,\eta_F L_\mathbf{BC}},
\end{equation}
with the characteristic exponent $\eta_F$.
We also study the exponent $\eta_F$ and observe that it increases linearly with temperature (see Figure \ref{fig:fidelityvsboundary2}$.\text{(d)}$).
 In fact, we obtain that $\eta_F = \alpha\, T + \beta$ with $\alpha \approx 0.970 \pm 0.005$ and $\eta_F\in[0,1.3]$ for $T\in[5.6,7]$.
Exponents were computed by fitting data in the interval $[L_\mathbf{BC}^{\text{min}},L_\mathbf{BC}^{\text{max}}]$, where $L_\mathbf{BC}^{\text{max}}=\min (2\,L_0/3,L_{\text{cut}})$ with $L_{\text{cut}}=\{L \in \mathbb R : \text{F}(L)\approx 1-10^{-14}\}$, and $L_\mathbf{BC}^{\text{min}}=\max (L_\mathbf{BC}^{\text{max}}-2\,L_0/5,6)$. The values of these parameters for the computation of the exponents was motivated by the fact that they define the largest range in which the curves show a clear exponential behavior.

\begin{figure}
\centering
  \begin{subfigure}[t]{0.48\textwidth}
  \includegraphics[width=0.95\textwidth]{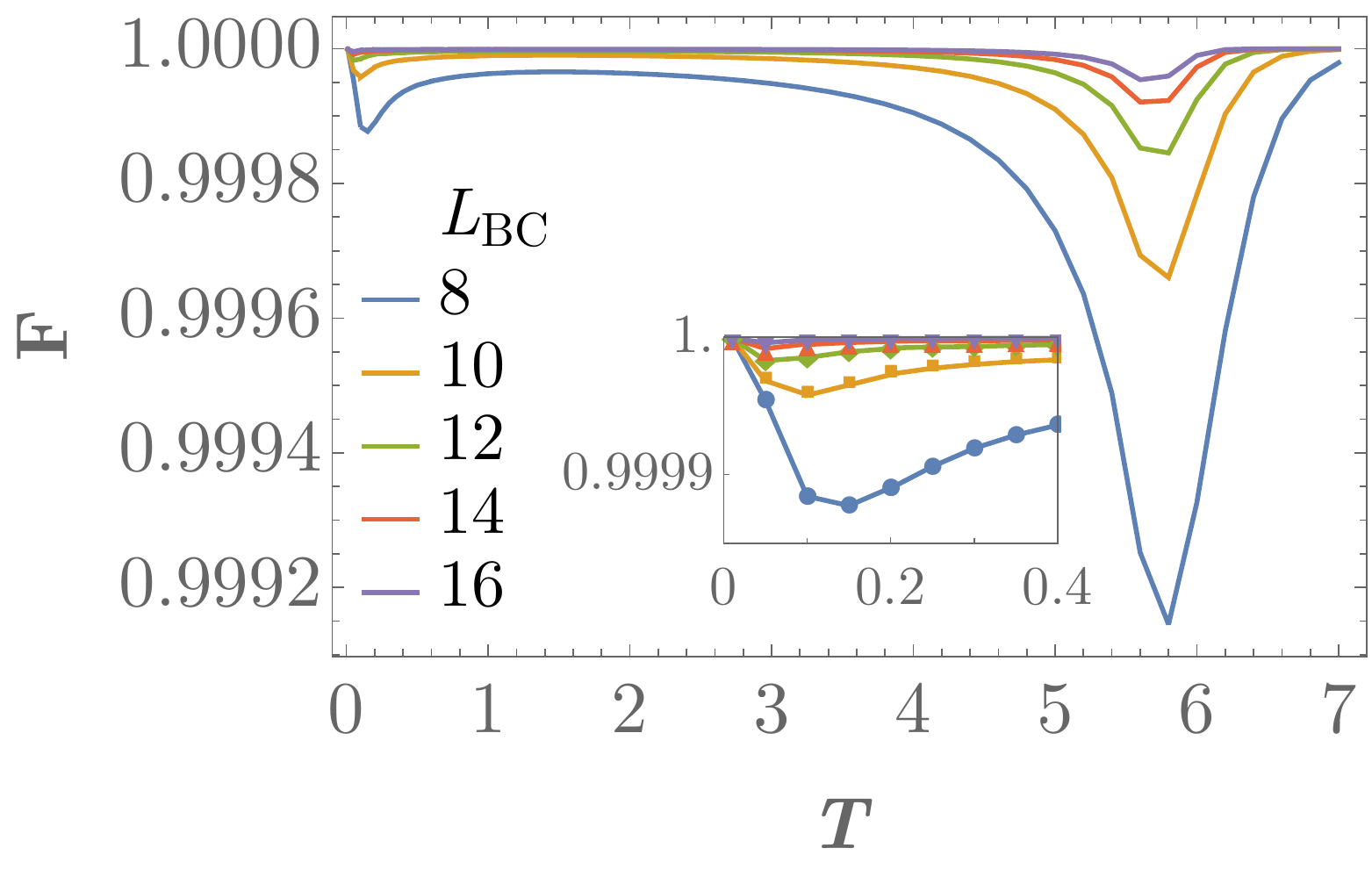}
  \caption{Fidelity vs temperature for different lengths $L_\mathbf{BC}=8,10,12,14,16$.
  Inset: Plot for data at temperature $T\leq 1$.
  Notice that the scaling of the plot is logarithmic, and the scaling of the inset is double logarithmic.}
  \end{subfigure}\hfill
  \begin{subfigure}[t]{0.48\textwidth}
  \includegraphics[width=0.9\textwidth]{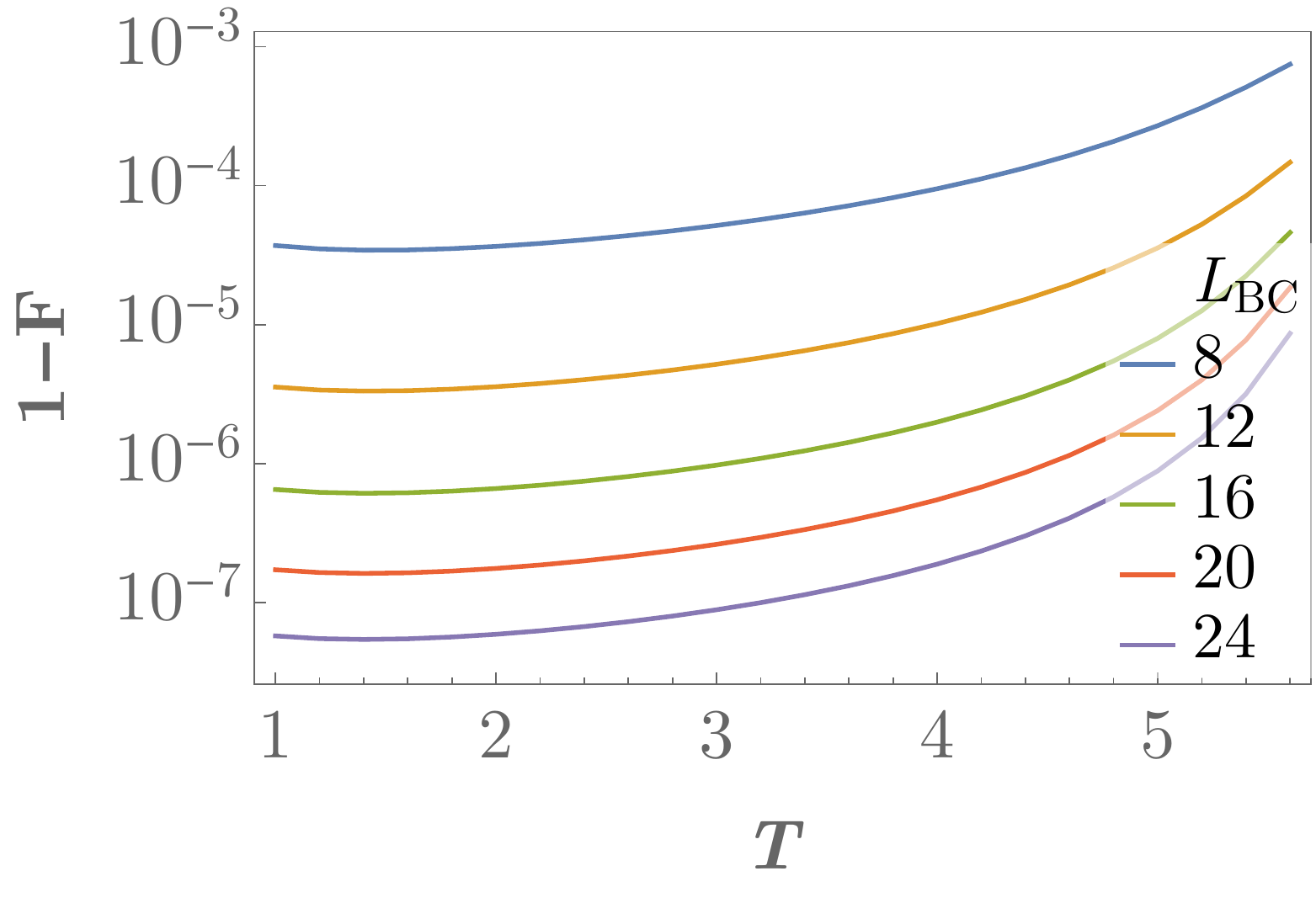}
  \caption{Fidelity error, $1-\fidelity(\rho_\mathbf{C},\rho_\mathbf{C'})$, vs temperature for $L_0=100$ and for length $L_\mathbf{BC}=8,12,16,20,24$.}
  \end{subfigure}\\
  \begin{subfigure}[t]{0.48\textwidth}
  \includegraphics[width=0.95\textwidth]{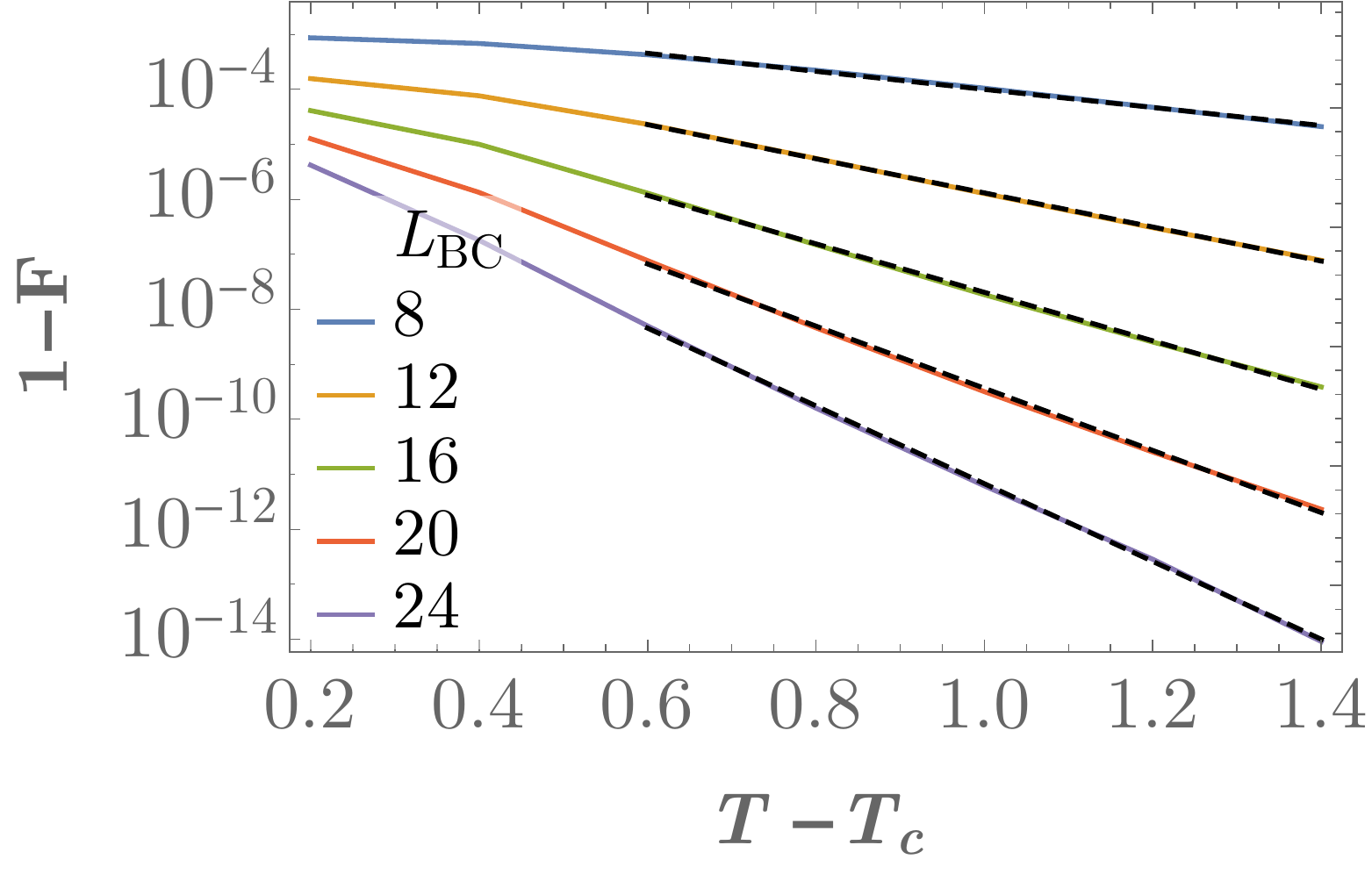}
  \caption{Fidelity error, $1-\fidelity(\rho_\mathbf{C},\rho_\mathbf{C'})$, vs temperature difference $T-T_c$ for $L_0=100$ and for length $L_\mathbf{BC}=8,12,16,20,24$.
  The dashed lines correspond to an exponential fit for $T\in[6.2,7]$}
  \end{subfigure}\hfill
  \begin{subfigure}[t]{0.48\textwidth}
  \includegraphics[width=0.9\textwidth]{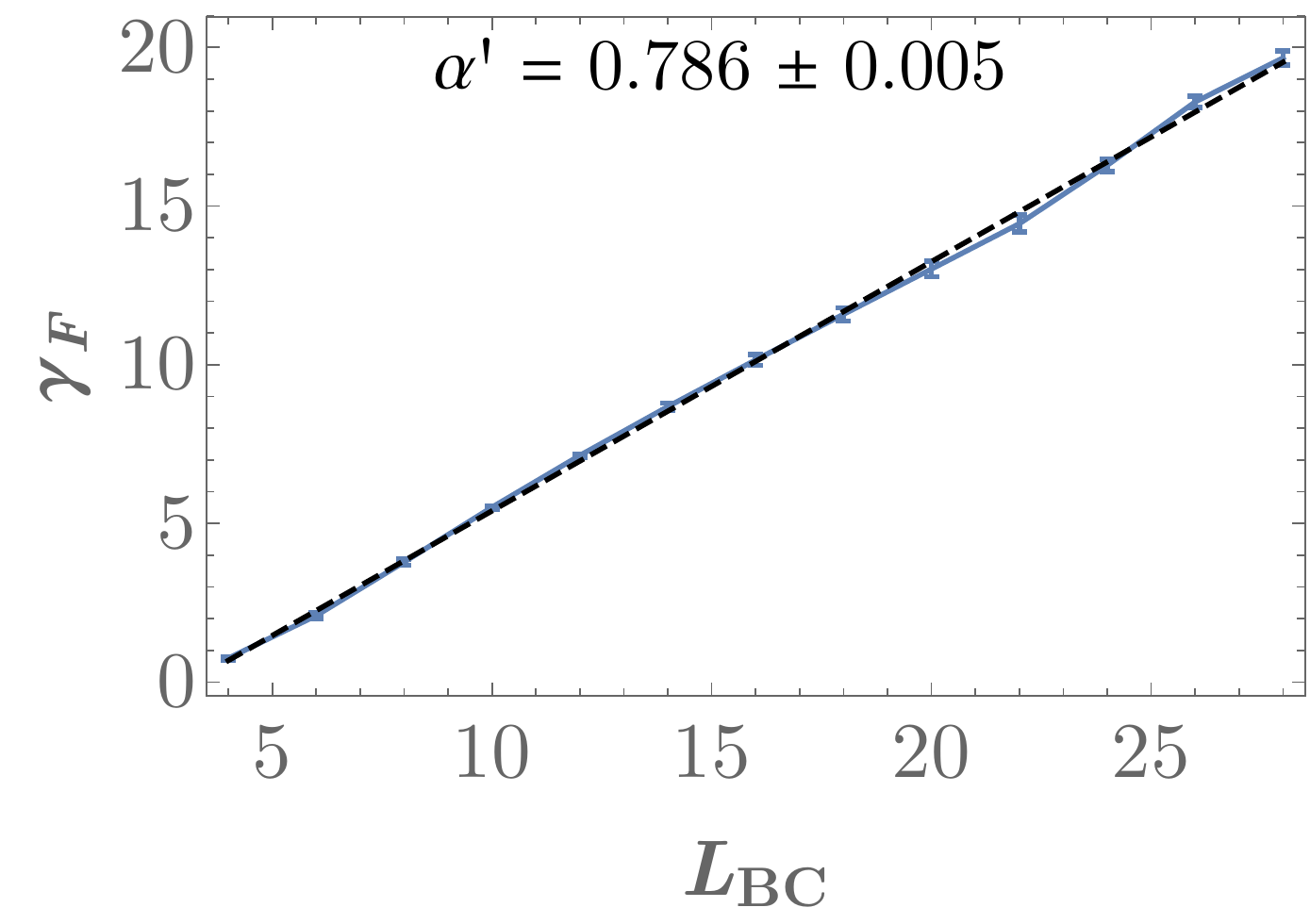}
  \caption{Exponent $\gamma_F$ of exponential fit vs the length of reference system size $L_\mathbf{BC}$.
  The dashed line corresponds to a linear fit with exponent $\alpha'$.
  Error uncertainty and error bars represent the standard error of the parameters, as obtained from the linear interpolations.}
  \end{subfigure}
\caption{\label{fig:fidelityvstemperature} Temperature dependence of the fidelity }
\end{figure}

We also investigate the behavior of the fidelity as a function of the temperature $T$ for different lengths $L_\mathbf{BC}$ (see Figure \ref{fig:fidelityvstemperature}$.\text{(a)}$).
We observe a global minimum at the critical temperature $T_c$ for any length $L_\mathbf{BC}$.
Unfortunately, we are not able to identify the behavior of the fidelity around the critical point at $T_c$. In fact, below the critical temperature, we do not observe a clear scaling (see Figure \ref{fig:fidelityvstemperature}$.\text{(b)}$), as the data could be equally fitted to an exponential or power-law function around the critical point $T_c$.
The same occurs for temperatures that are just above the critical temperature, $T_c$.
However,  at very large temperatures, $T>>T_c$, it is possible to see that the fidelity goes exponentially to $1$  (see Figure \ref{fig:fidelityvstemperature}$.\text{(c)}$), that is,
\begin{equation*}
  1-\fidelity(\rho_\mathbf{C},\rho_\mathbf{C'}) \propto \e^{-\,\gamma_F\,T}
\end{equation*}
with exponent $\gamma_F$.
We also analyze how this exponent depends on the system size and obtain that $\gamma_F \propto \alpha' L_{BC}+\beta'$ with a factor $\alpha' \approx 0.786 \pm 0.005$ (see Figure \ref{fig:fidelityvstemperature}$.\text{(d)}$). The fits have been obtained for data with $T\in[6.2,7]$.

Additionally, we observe the presence of a local minimum at low temperatures.
We have verified the existence of this minimum by computing the fidelity using a different approach. We constructed an approximation to the states at low temperature by computing their elements in the Fock basis and neglecting all the elements with a large number of bosons. We observe that for large numbers the truncation error seems to be negligible and obtain the same results for the fidelity.
We have also observed that the ratio between this low-temperature minimum and the one at the critical temperature decreases when the system size increases, indicating that this may be a finite-size effect.

\begin{figure}[t!]
\centering
\includegraphics[width=0.5\textwidth]{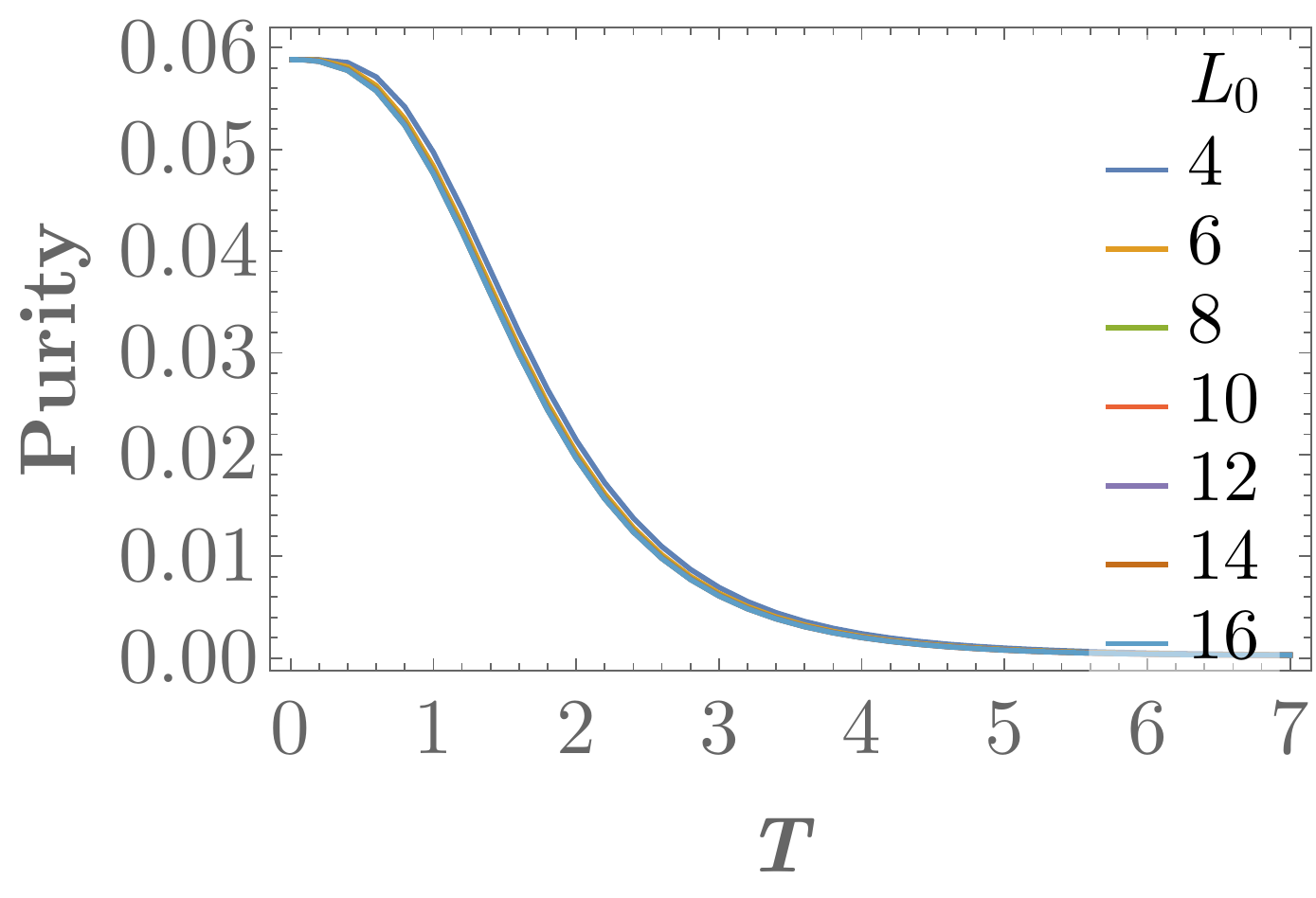}
\caption{\label{fig:purity}Purity of the partial state $\rho_\mathbf{C}$ vs temperature $T$ for different values of system size $L_0=4,6,8,10,12,14,16$.}
\end{figure}

As a last remark, it is necessary to highlight that the fidelity is extremely high, with values $F>0.988$ for any given parameters.
To understand why this happens, let us study the purity, a measure of how much a given state $\rho$ is mixed and that it is given by
\begin{equation*}
  \purity(\rho) = \tr(\rho^2).
\end{equation*}
The purity satisfies $\frac{1}{d}\leq\purity\leq 1$ : $\purity=1$ implies that the state $\rho$ is pure and $\purity=d^{-1}$ means that it is completely mixed, where $d$ is the Hilbert space dimension.
We compute the purity of the partial state $\rho_\mathbf{C}$ for different system sizes $L_0$ (see Figure \ref{fig:purity}) and obtain that the purity $\purity<0.06$ for any system size $L_0\in[4,16]$.
We therefore see that already for these small system sizes, $L_0\leq 16$, the reduced state $\rho_\mathbf{C}$ is extremely mixed and the computed purity is almost independent of the system size, which may explain why we are obtaining such high values of the fidelity.

\section{Relation to correlations}
\label{sec:relcorrelations}
In this section, we compare the decay of the distinguishability $1-F(\rho_C,\rho_{C'})$ to that of the density-density correlations in the finite-size model \eqref{eq:BEmodel}.
Let us first review how density-density correlations,
\begin{equation*}
\corr(n_{\textbf{i}},n_{\textbf{j}}) := \expect{n_i n_j}-\expect{n_i} \expect{n_j}
\end{equation*}
with density operator $n_i:=b_i^\dagger b_i$, decay in this model \cite{Pitaevskii2016}.

\begin{figure}[t!]
\centering
\begin{subfigure}[t]{0.48\textwidth}
  \includegraphics[scale=0.45]{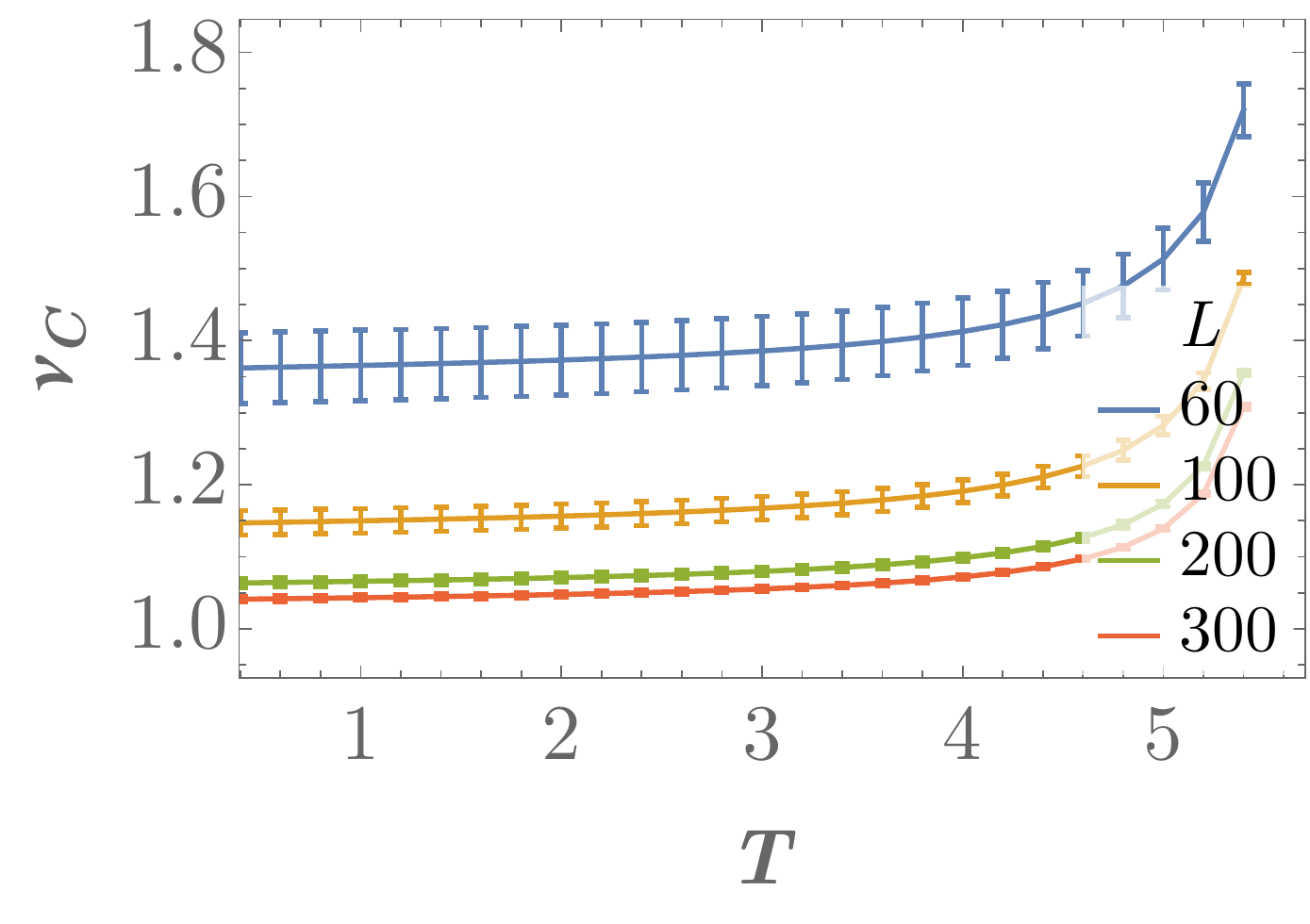}
  \caption{Exponents for the power-law fits for $T\leq T_c$.}
\end{subfigure}\hfill
\begin{subfigure}[t]{0.48\textwidth}
  \includegraphics[scale=0.46]{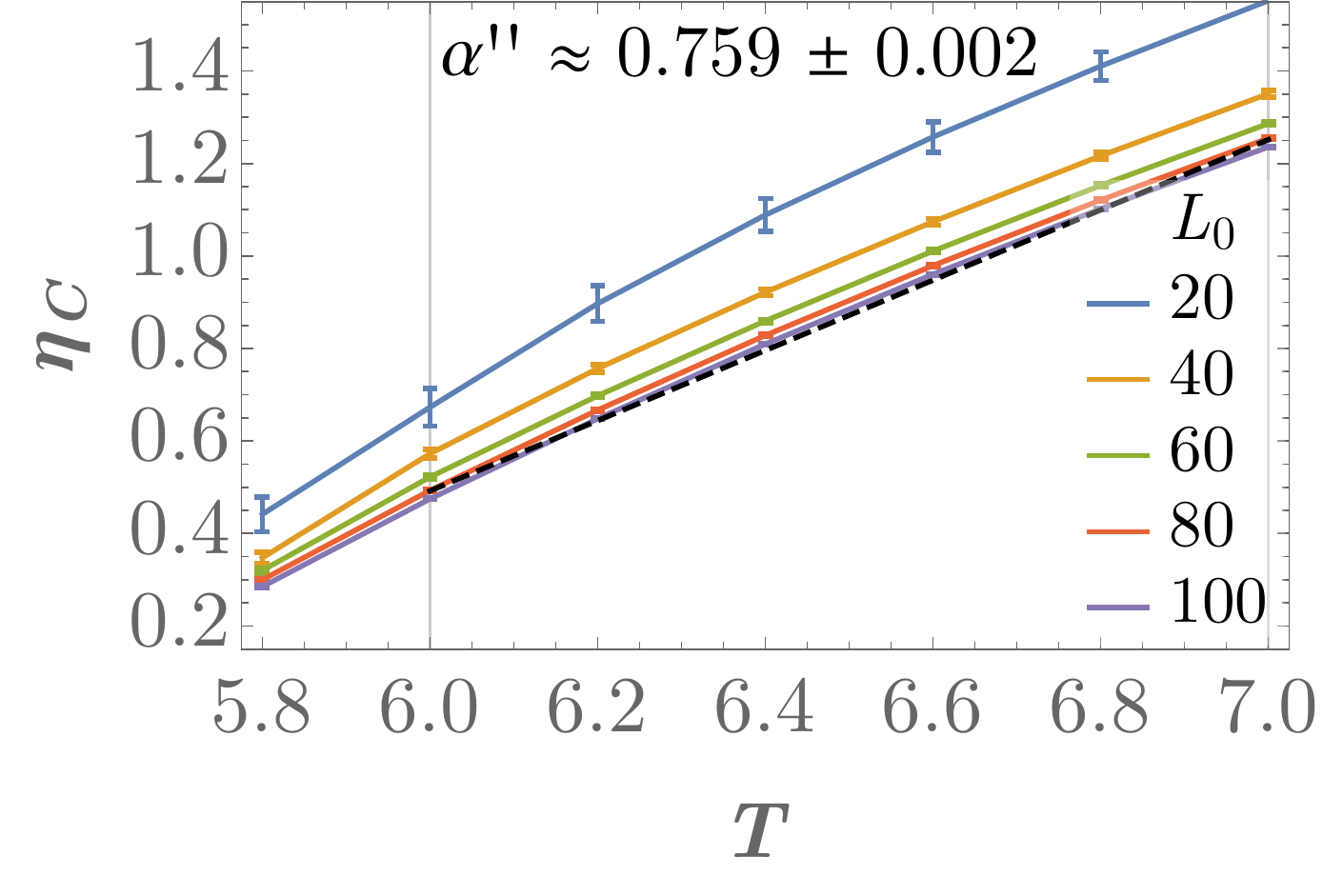}
  \caption{Exponents for the exponential fits for $T>T_c$.
  The dashed line corresponds to a linear fit with exponent $\alpha''$. }
\end{subfigure}
\caption{\label{fig:exponents1} Exponents of the correlation decay above and below the critical temperature. Error uncertainty and error bars represent the standard error of the parameters, as obtained from fits (more details in \ref{app:correlations}). }
\end{figure}

Using Wick's theorem we obtain that these correlations can be rewritten as
\begin{equation*}
  \corr(n_{\textbf{i}},n_{\textbf{j}}) = \expect{b_\textbf{i}^\dagger\,b_\textbf{j}}\,\expect{b_\textbf{i}\,b_\textbf{j}^\dagger}-\expect{b_\textbf{i}^\dagger\,b_\textbf{j}^\dagger}\,\expect{b_\textbf{i}\,b_\textbf{j}} ,
\end{equation*}
and thus their behavior is fully determined by the elements of the covariance matrix. These in fact decay exponentially with the distance $\dist:=|\textbf{i}-\textbf{j}|$ above the critical temperature  $T_c$, thus in this regime
\begin{equation*}
  \corr(n_{\textbf{i}},n_{\textbf{j}}) \propto \e^{-\,\eta_C\dist}.
\end{equation*}
Below $T_c$ the elements of the covariance matrix decay polynomially to a constant and, thus, the correlations also show a polynomial decay to a constant $\corr^{\infty}$:
\begin{equation*}
  \corr(n_{\textbf{i}},n_{\textbf{j}}) - \corr^{\infty} \propto \frac{1}{\dist^{\nu_C}},
\end{equation*}
where $\corr^{\infty}=\lim_{\dist \longrightarrow \infty}\corr(n_{\textbf{i}},n_{\textbf{j}}) = n_0^2$ \cite{Pitaevskii2016}.\\
We have numerically determined the exact values of $\nu_C$ and $\eta_C$ by considering correlations between the sites $\textbf{i}=(i,0,0)$ and $\textbf{j}=(j,0,0)$ (for details, see \ref{app:correlations}).
We found that the exponent $\nu_C$ is weakly dependent on temperature as $\nu_C \approx 1.05 \pm 0.01$ for $T\leq 4$ and monotonically increases with the temperature for $T\in [4,5.6]$ for $L_0=300$ (see Figure \ref{fig:exponents1}$.\text{(a)}$). It is worth noting here that for the ideal Bose gas, this exponent is equal to $1$~\cite{Pitaevskii2016} and our numerical estimates get close to it (recall that the error only refers to the linear interpolation).
The exponent $\eta_C$ increases linearly with the temperature $T$ as $\eta_C=\alpha'' T+\beta''$ with a factor $\alpha''\approx 0.759\pm 0.002$ and with values $\eta_C\in[0,1.2]$ for  $L_0>80$ (see Figure \ref{fig:exponents1}$.\text{(b)}$).

Given these results, we observe that the distinguishability $(1-F)$ and the correlations show the same behavior and both decay exponentially above the critical temperature.
Below the critical temperature, the distinguishability decays polynomially to zero. This is not surprising, as the distinguishability has to be zero when $L_{BC}=L_0$, but it is interesting that the decay remains polynomial despite the long-range correlations present in the system, namely $\corr^{\infty}= n_0^2>0$.

Let us finally compare the temperature dependence of the exponents. Below and at the critical temperature, the exponents $\nu_F$ and $\nu_C$ are constant for $T\leq 4$, but $\nu_F$ grows, while $\eta_F$ decreases with temperature.
Above the critical temperature, $\eta_F$ and $\eta_C$ both increase linearly with the temperature.

\section{Conclusions}
In this work, we study to what extent temperature can be defined locally for a three-dimensional discrete version of the Bose-Einstein model \eqref{eq:BEmodel} at the grand canonical state \eqref{eq:GCstate} with fixed particle density $n$.
This model undergoes a non-zero-temperature phase transition in a similar fashion as the Bose-Einstein model, i.e., the system condensates to the ground state below a critical temperature $T_c$.

We obtain that the reduced density of the grand canonical state to a given region converges as we consider systems of larger and larger sizes. The convergence rate is exponential in the system size above the critical temperature and polynomial below.
This rapid convergence then means that temperature can be defined locally: if temperature is defined through the grand canonical ensemble, a very good approximation to the reduced state of the full system can be obtained by considering another subsystem slightly larger than the subsystem of interest.
The quality of the local temperature description attains a minimum at $T=T_c$ at any boundary size $L_\mathbf{BC}$, which is a signature of the phase transition. For all the studied situations, however, the reported fidelities between the actual state and the effective local grand canonical state are very high for any boundary length and temperature. We suspect that this is due to the fact that the reduced states are always highly mixed and therefore almost independent of the system size $L_{BC}$. We provide evidence for this by showing that the purity of the reduces states is very low.
As one associates low correlations with locality, we compared our results to the decay of correlations in this model. We studied density-density correlations and observe an equivalence between the qualitative behaviors of the locality and the correlations, as they behave polynomially for $T\leq T_c$ and exponentially for $T>T_c$.

Prior to our work, there were several results demonstrating the validity of a local temperature description in systems with local interactions. In particular, this was proven for one-dimensional spin systems in~\cite{1367-2630-17-8-085007} using the fact that correlations decay exponentially for any value of $T>0$. Also, in~\cite{Kliesch2014}, the quality of approximating the state of a subsystem in an equilibrium state by the thermal state of a slightly bigger system was bounded by a function depending on the correlations in the thermal state. None of these works, however, was able to provide information about how the local temperature description applies in the presence of non-zero temperature phase transitions. Our results show, admittedly in a rather simple example, that local temperatures may also be assigned in this regime.

\section*{Acknowledgements}
We acknowledge financial support from the European Union's Marie Sk\l{}odowska-Curie Individual Fellowships (IF-EF) programme under GA: 700140, the European Union's Horizon 2020 program through the ERC StG WASCOSYS (No. 636201), the ERC CoG GAPS (No. 648913), the ERC AdG CERQUTE, the ERC AdG QENOCOBA (No. 742102), the AXA Chair in Quantum Information Science, the Government of Spain (FIS2020-TRANQI and Severo Ochoa CEX2019-000910-S and SEV-2015-0554), Fundacio Cellex and Fundacio Mir-Puig, Generalitat de Catalunya (SGR 1381, QuantumCAT and CERCA Programme) and  from the DFG (German Research Foundation) under Germanys Excellence Strategy (EXC-2111-390814868).

\section*{References}
\bibliographystyle{iopart-num}
\bibliography{loctemfin}

\newpage
\appendix

\section{Phase transition}\label{app:phase_transition}
\begin{figure}
\centering
\begin{subfigure}[t]{0.48\textwidth}
  \includegraphics[width=0.93\textwidth]{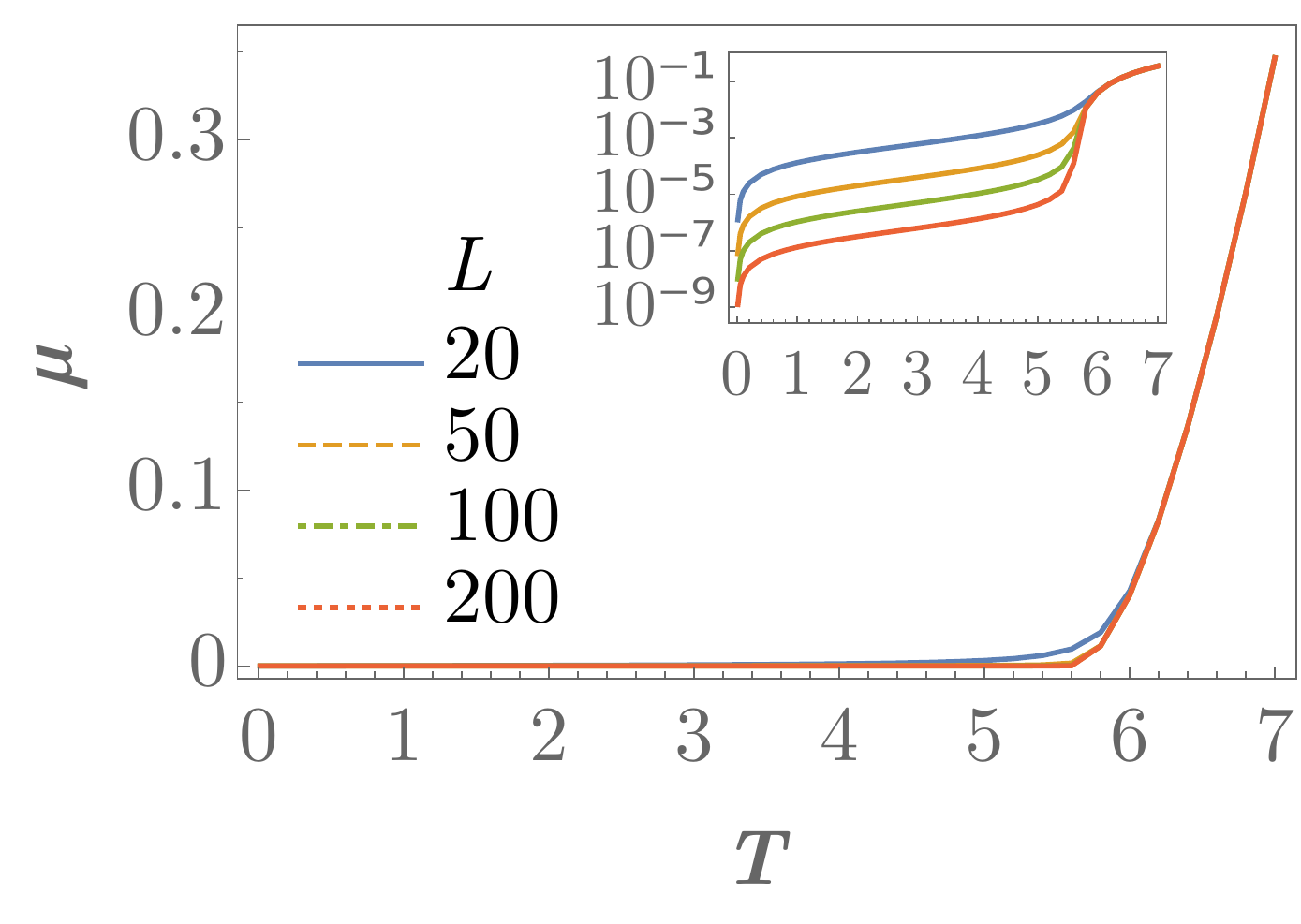}
  \caption{Chemical potential, $\mu$, vs temperature, $T$, for density  $n=1$.}
\end{subfigure}\hfill
\begin{subfigure}[t]{0.48\textwidth}
  \includegraphics[width=0.93\textwidth]{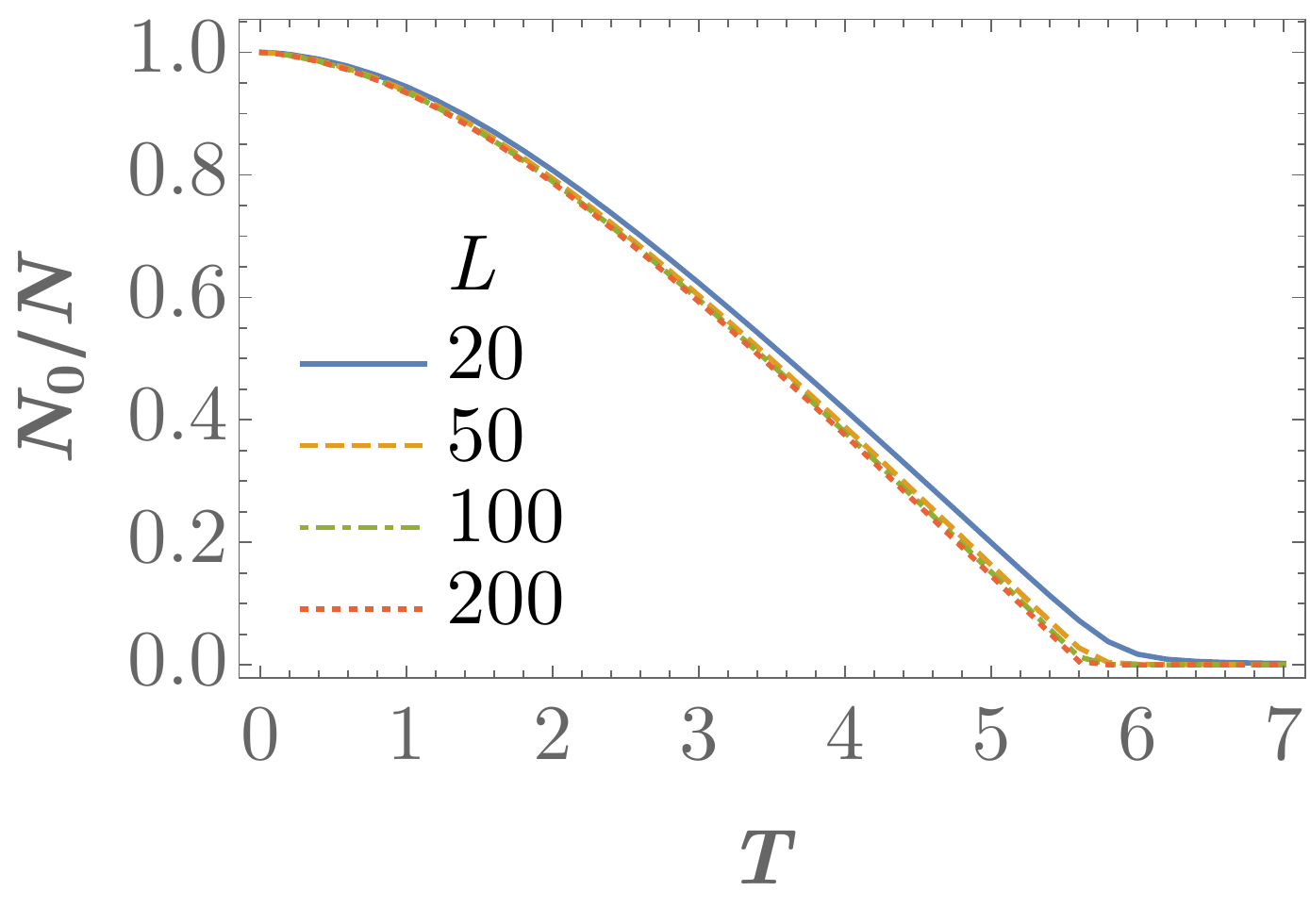}
  \caption{Zero-momentum density, $N_0/N$, vs temperature, $T$, for density $n=1$ and different system sizes.}
\end{subfigure}\\
\begin{subfigure}[t]{0.48\textwidth}
  \includegraphics[width=0.96\textwidth]{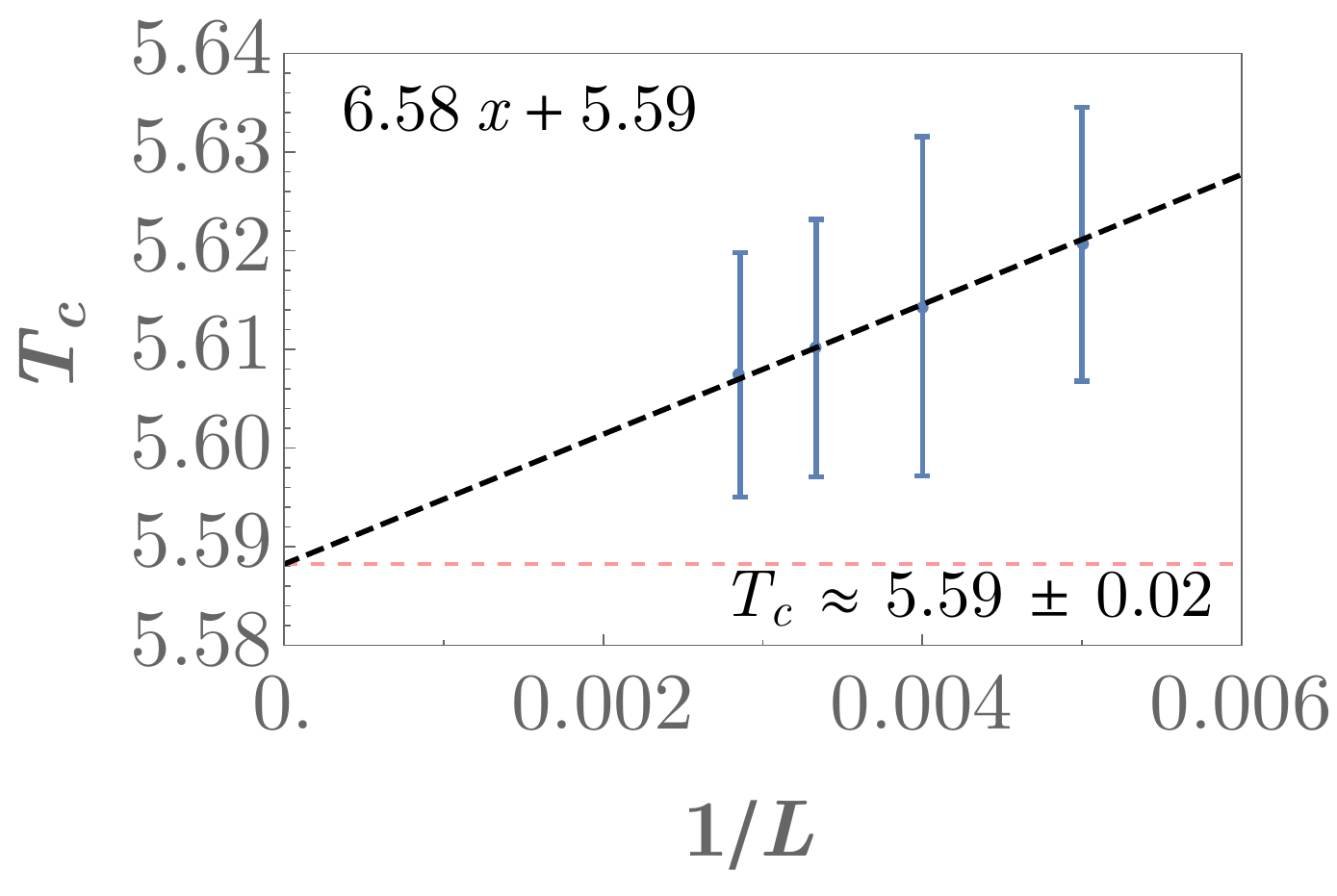}
  \caption{Critical temperature, $T_c$ vs inverse distance, $1/L$, for density $n=1$.
  Data involves system sizes $L=100,200,250,300,350$.
  The red dashed line highlights the critical temperature at the thermodynamic limit $T_c$.
  Error uncertainty and error bars represent the standard error of the parameters, as obtained from fits.}
\end{subfigure}\hfill
\begin{subfigure}[t]{0.48\textwidth}
  \includegraphics[width=0.98\textwidth]{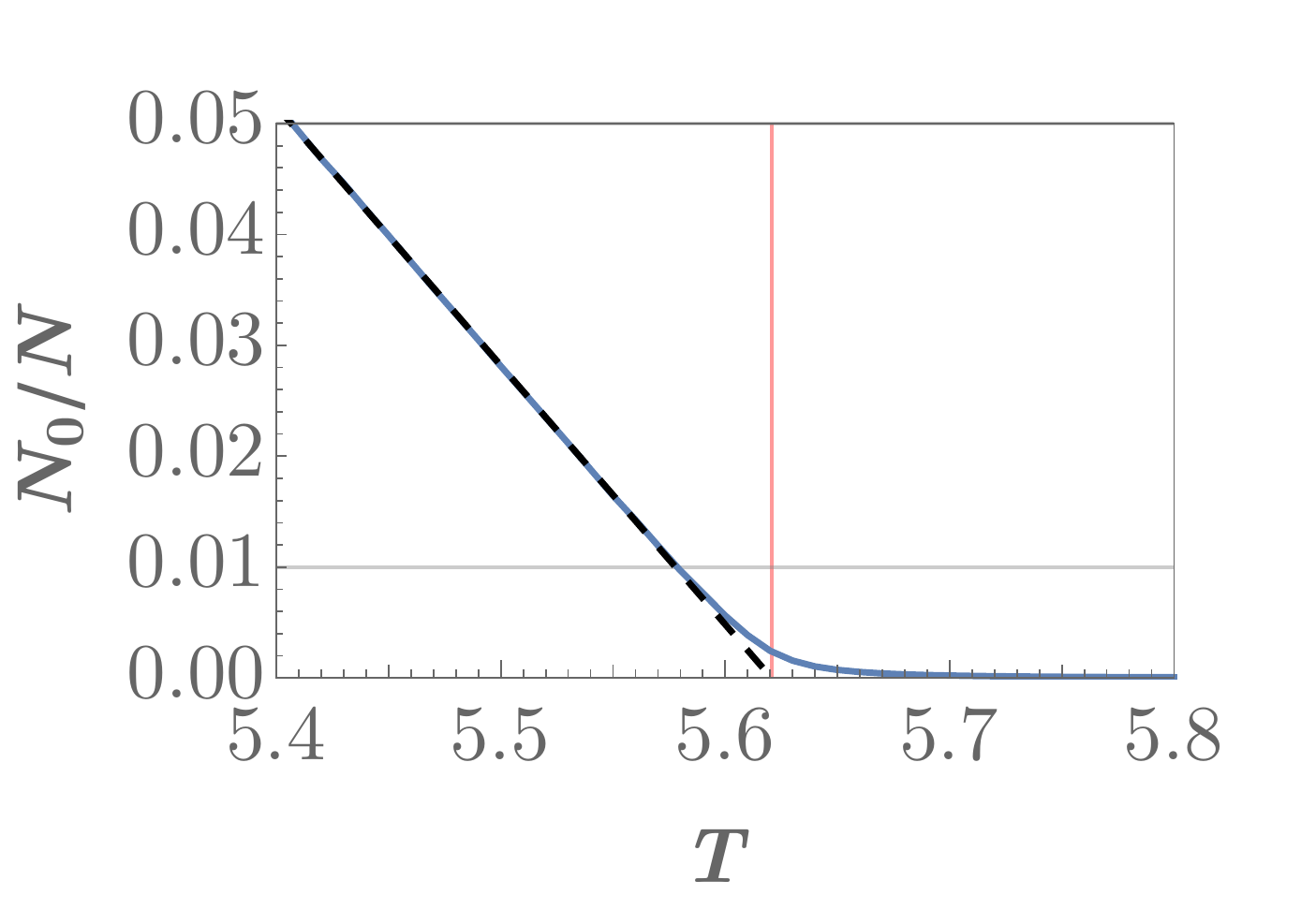}
  \caption{Zero-momentum density, $N_0/N$, vs temperature, $T$, for density $n=1$ and system size $L=200$.
  Data and fit function are represented by a solid and a dashed line, respectively, while the red line highlights the estimated critical temperature $T_c$ for each system size.
  The linear fit has been obtained from data with values between $N_0/N \in (0.01,0.05)$ (the min. value is represented by a grey line).
  The critical temperature $T_c$ for each length $L_0$ was obtained by fitting data between values $N_0/N \in [0.01,0.05]$.}
\end{subfigure}
\caption{\label{Num-Cond} Temperature dependence of the chemical potential and the zero-momentum density, and size dependence of the critical temperature. }
\end{figure}

In this section, we characterize the phase transition of the model.
First, we calculate the value of the chemical potential as a function of the temperature and system size, $\mu = \mu(\beta,L)$, such that the particle density is fixed at $n=1$. The obtained values of $\mu$ are plotted in Figure \ref{Num-Cond}$.\text{(a)}$.
We  then analyze the population ratio between ground-state particles and total number of particles, $N_0/N$, as a function of the temperature, $T$, for different system sizes, $L$ (see Figure \ref{Num-Cond}$.\text{(b)}$).
As expected, we observe a phase transition at critical temperature, $T_c$, and a condensate for any temperature $T\leq T_c$.
As the phase transition only makes sense at the thermodynamic limit, we obtain that the critical temperature $T_c \approx 5.59\pm 0.02$ by computing an estimation for each system size, $L$, and making a finite size analysis (see Figure \ref{Num-Cond}$.\text{(c)}$).
The estimation of the temperature for each system size was computed by linearly fitting data for $N_0/N \in [0.01,0.05]$ and extracting the temperature at which the linear fit goes to zero (see Figure \ref{Num-Cond}$.\text{(d)}$).
The error estimates correspond to the standard error, as obtained from the fits for each system size.
Notice that we assume a linear behavior of $T_c$ with the inverse system size $1/L$, which is a valid assumption for the given data set.

\section{Analysis of the fidelity, entropy and purity for different subsystems}

In this section, we study how the magnitudes of the fidelity, entropy and purity behave as a function of the temperature for different subsystems.
We consider subsystems consisting of 2 consecutive sites ($2\times1\times1$), 4 a plane with 4 sites ($2\times2\times1$) and a cube with 8 sites ($2\times2\times2$).

\begin{figure}[t!]
	\centering
  \begin{subfigure}[t]{\textwidth}
    \centering
    \begin{minipage}{0.3\textwidth}
      \includegraphics[width=\textwidth]{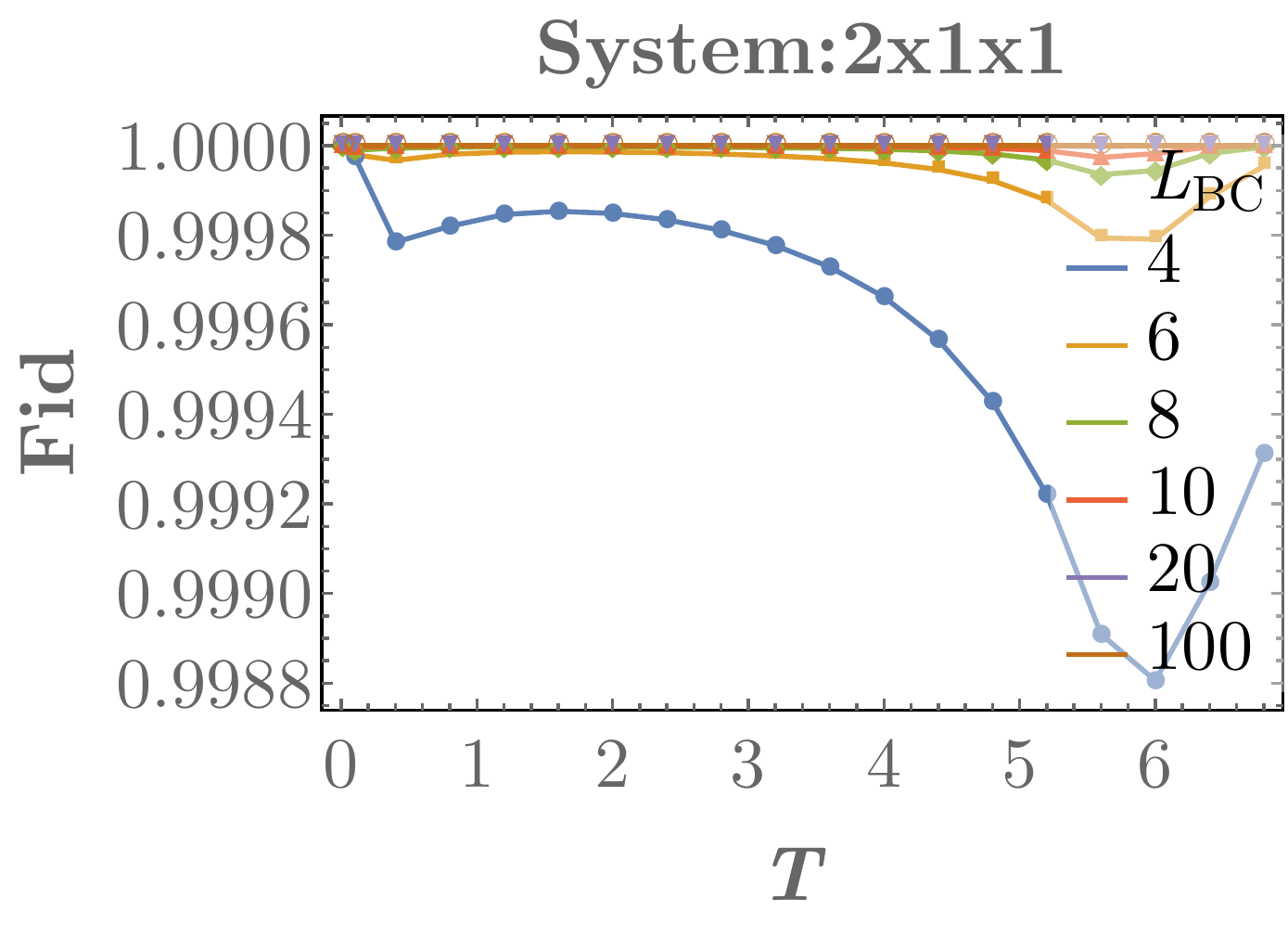}
    \end{minipage}
    \begin{minipage}{0.3\textwidth}
      \includegraphics[width=\textwidth]{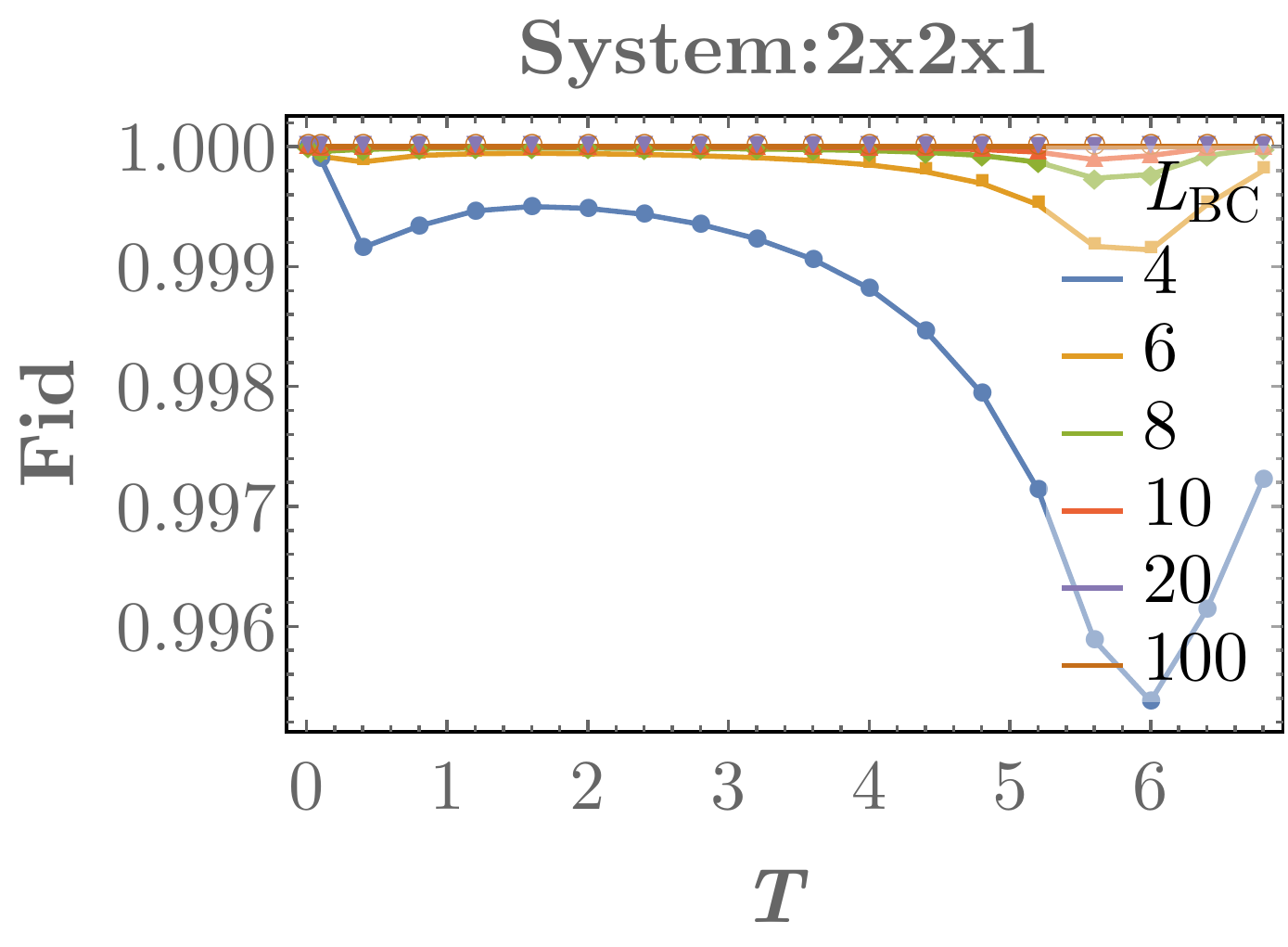}
    \end{minipage}
    \begin{minipage}{0.3\textwidth}
      \includegraphics[width=\textwidth]{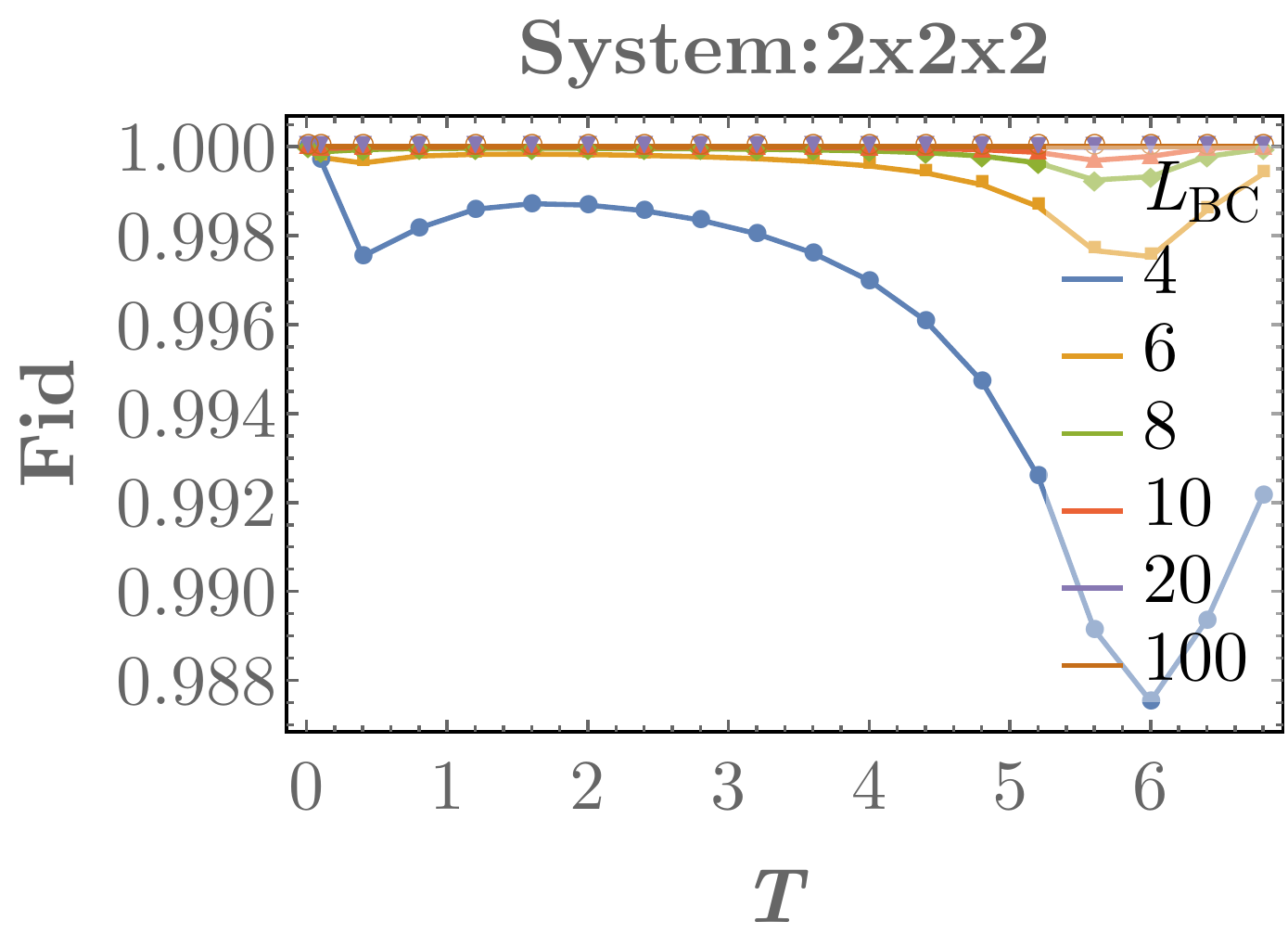}
    \end{minipage}
  \end{subfigure}\\
  \begin{subfigure}[t]{\textwidth}
    \centering
    \begin{minipage}{0.3\textwidth}
      \includegraphics[width=\textwidth]{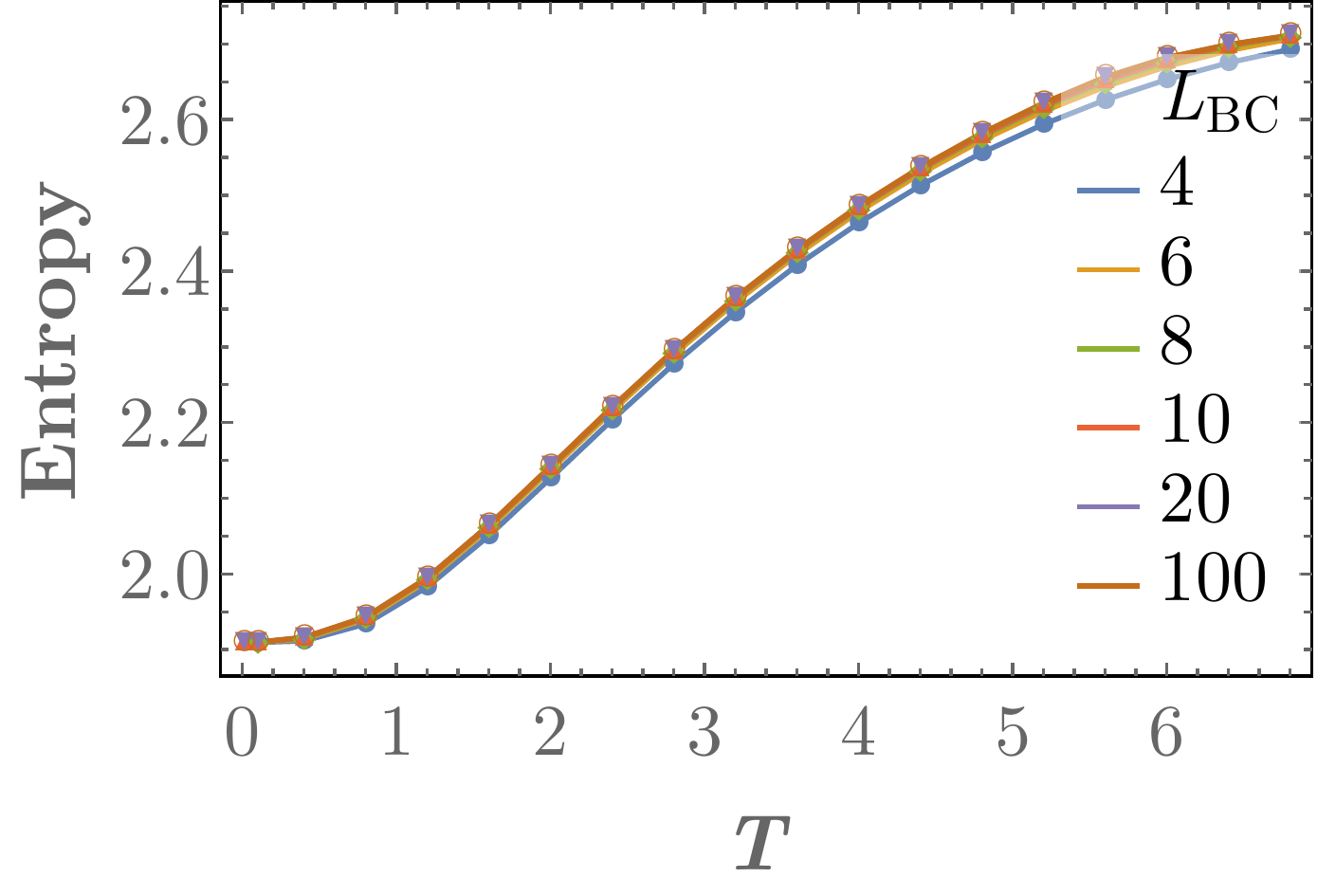}
    \end{minipage}
    \begin{minipage}{0.3\textwidth}
      \includegraphics[width=\textwidth]{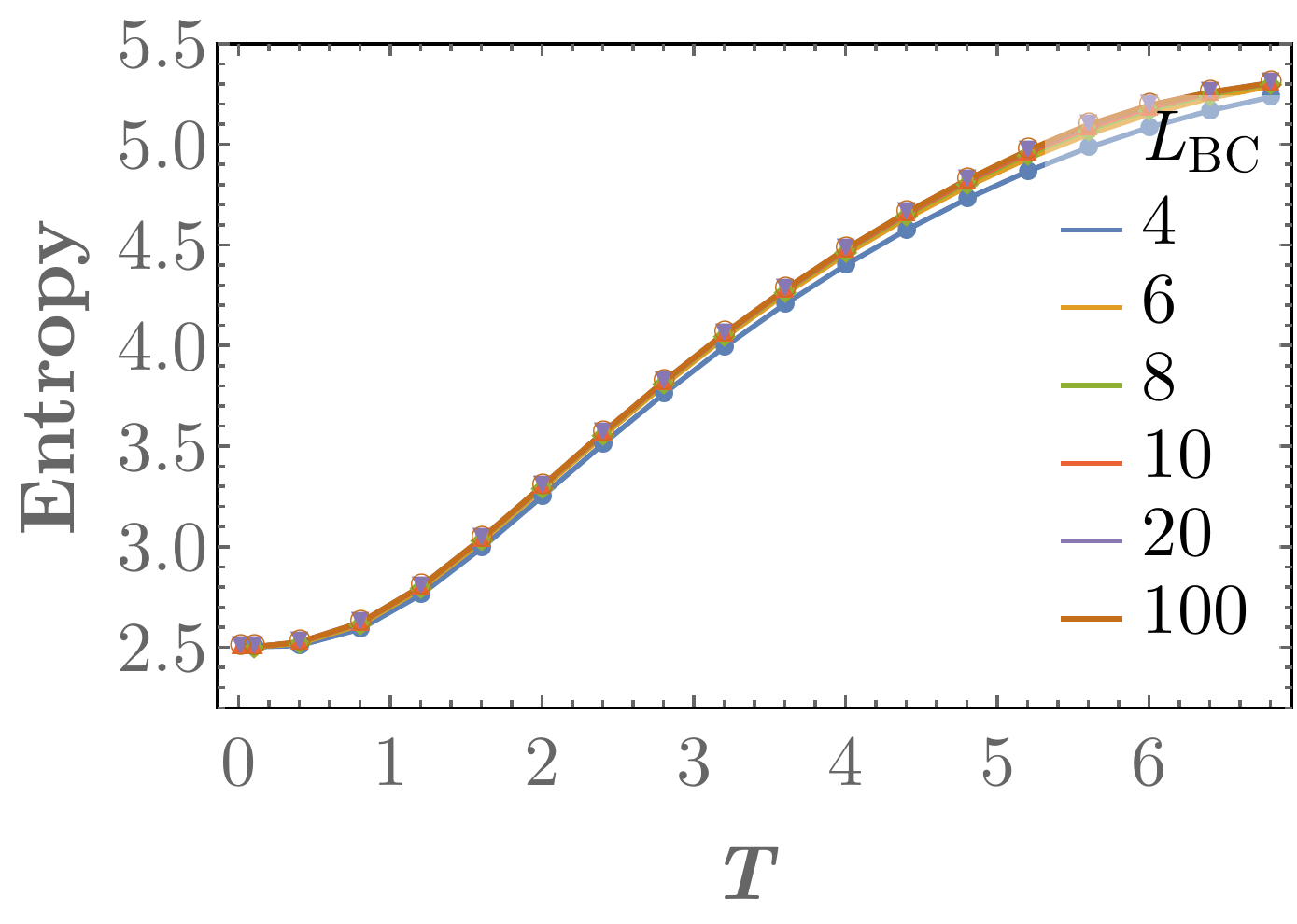}
    \end{minipage}
    \begin{minipage}{0.3\textwidth}
      \includegraphics[width=\textwidth]{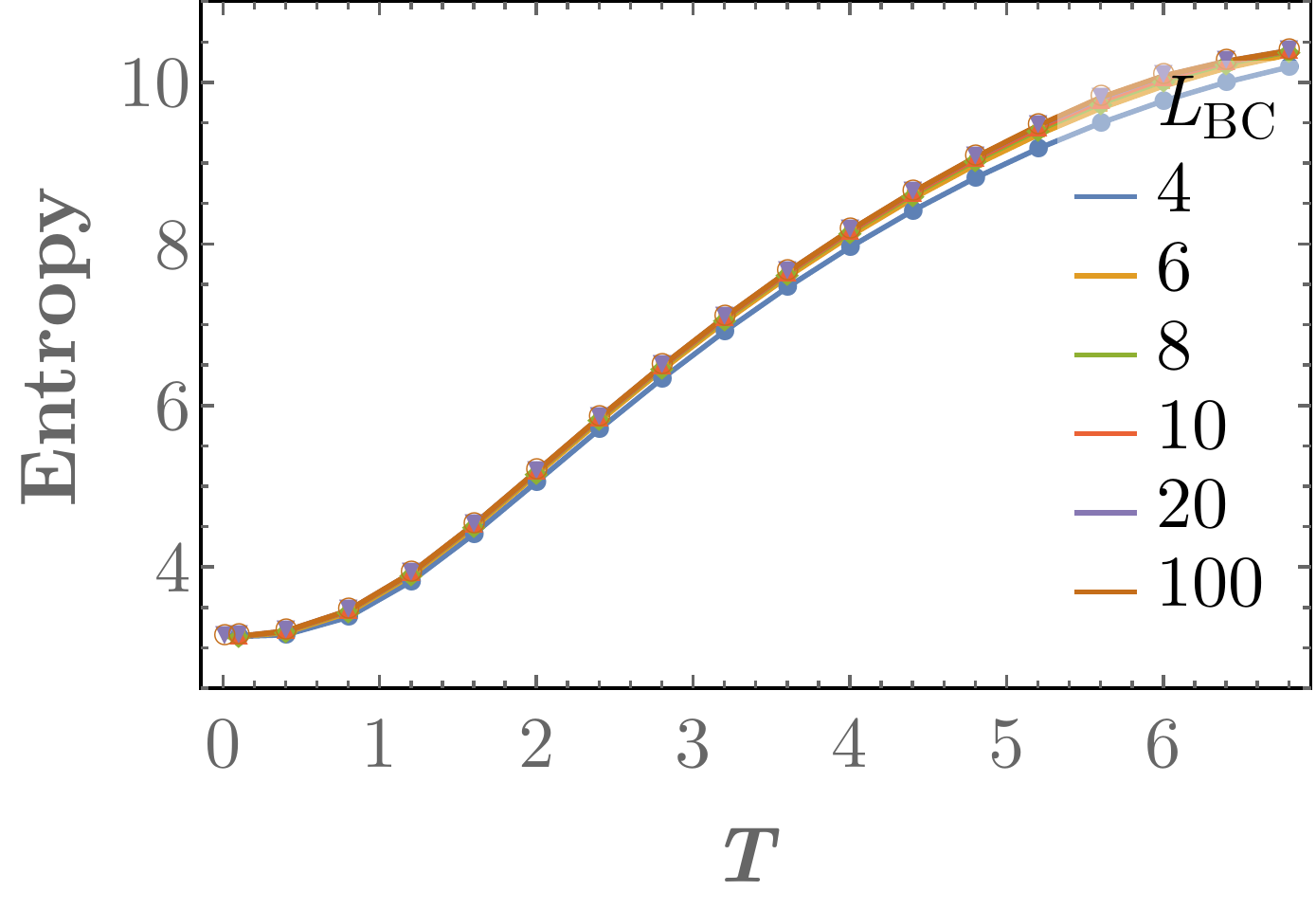}
    \end{minipage}
  \end{subfigure}\\
  \begin{subfigure}[t]{\textwidth}
    \centering
    \begin{minipage}{0.3\textwidth}
      \includegraphics[width=\textwidth]{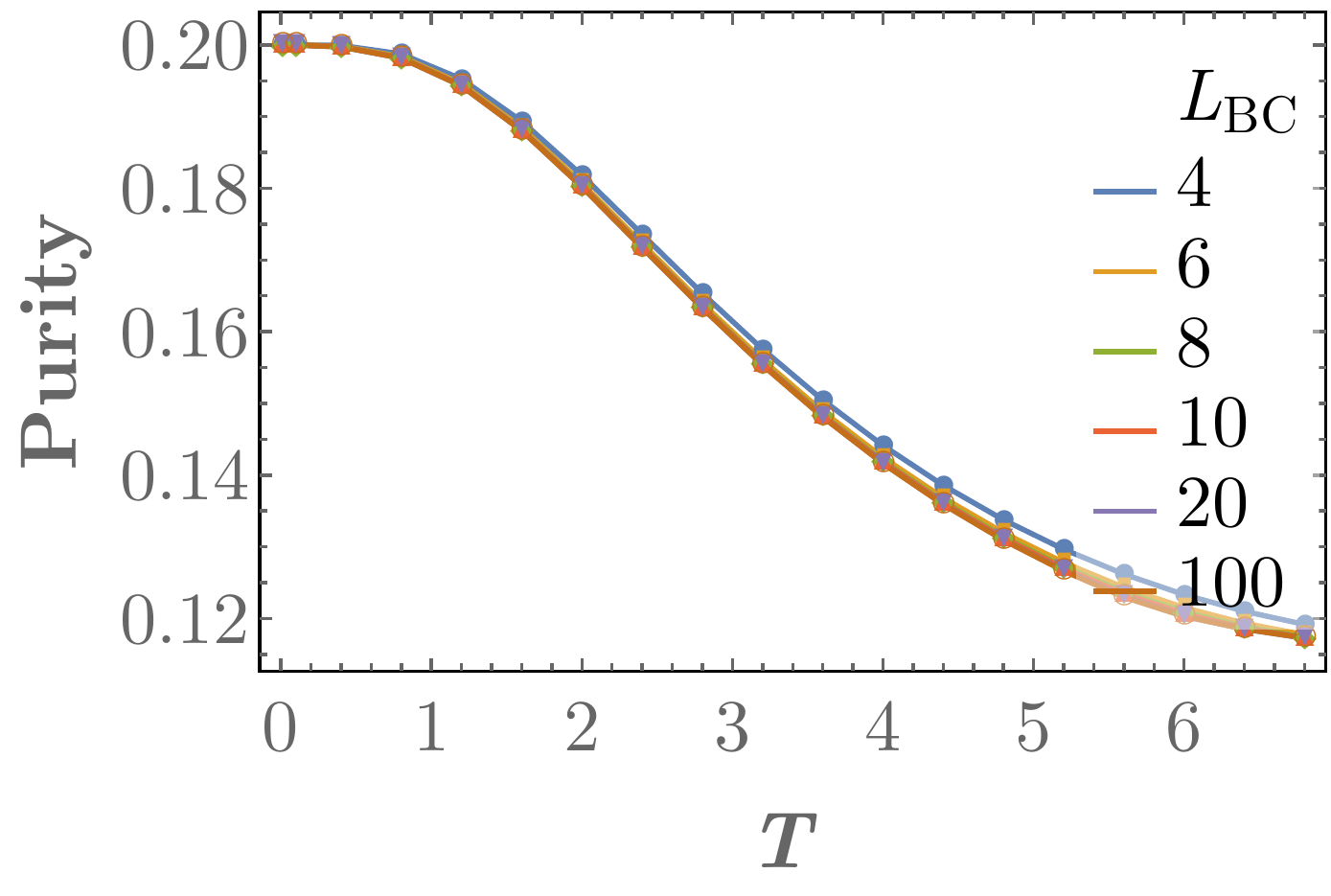}
    \end{minipage}
    \begin{minipage}{0.3\textwidth}
      \includegraphics[width=\textwidth]{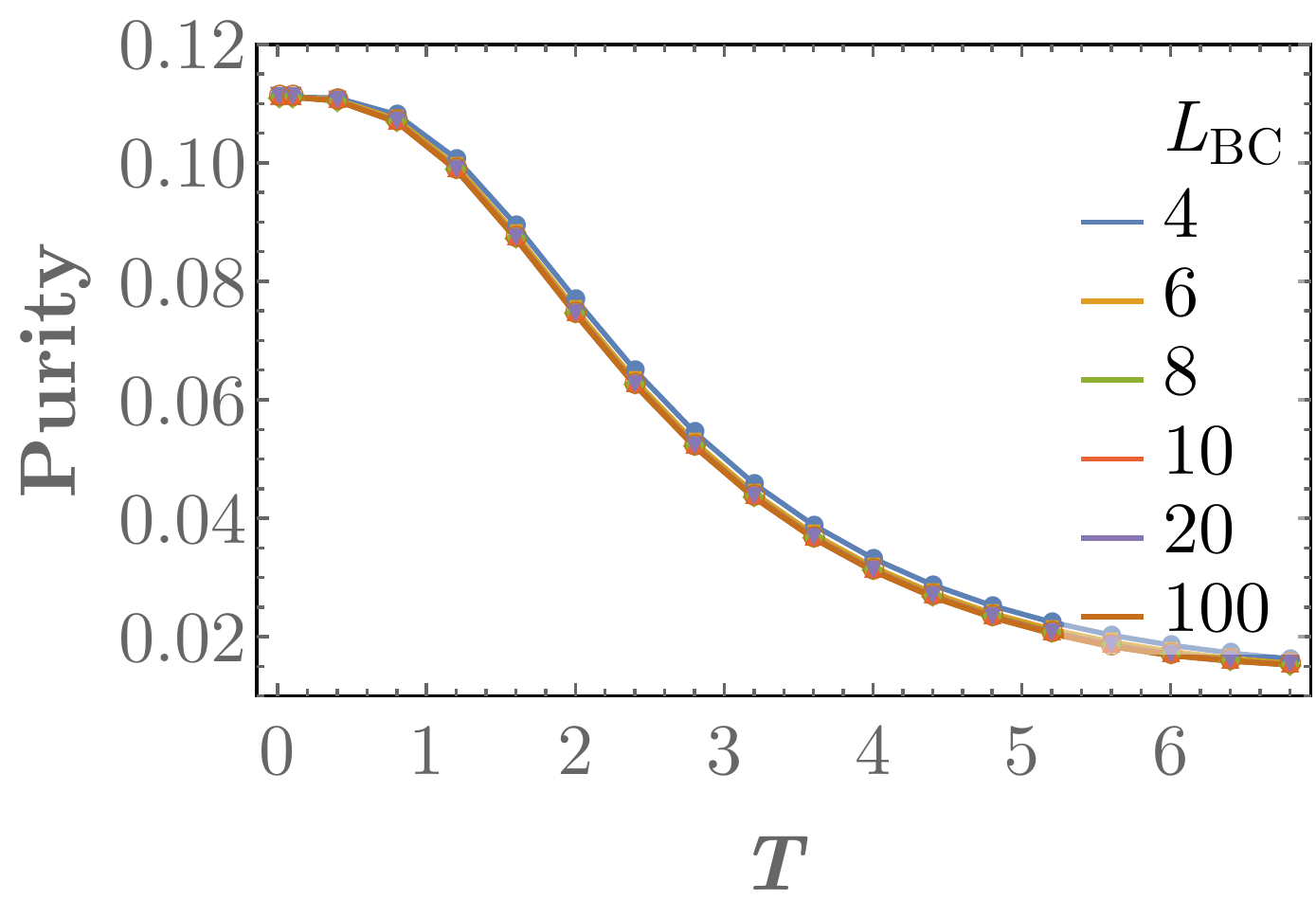}
    \end{minipage}
    \begin{minipage}{0.3\textwidth}
      \includegraphics[width=\textwidth]{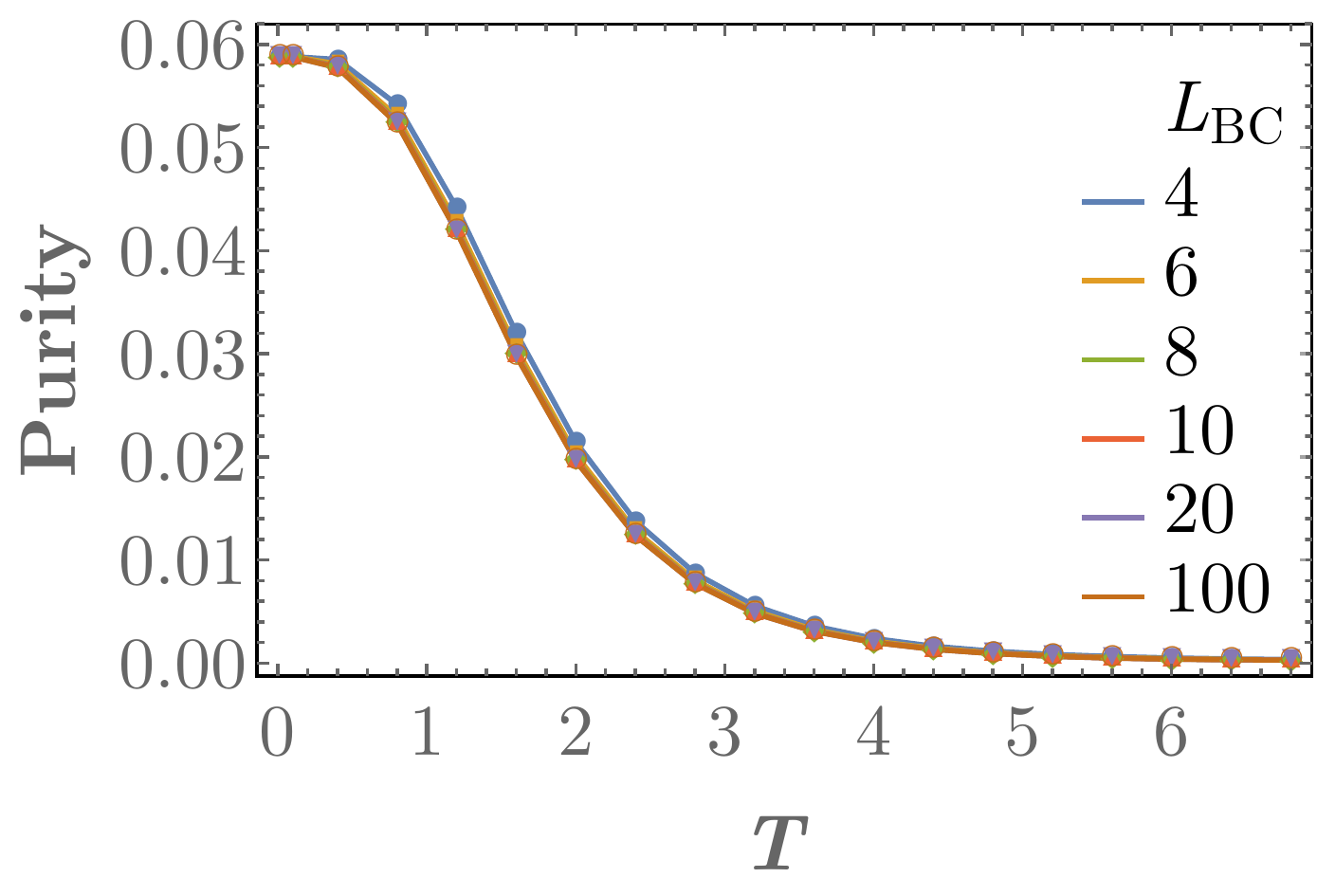}
    \end{minipage}
  \end{subfigure}
	\caption{\label{fig:fidelitystate2} Magnitudes of fidelity, entropy and purity for subsystems $2\times1\times1$ (left), $2\times2\times1$ (center) and $2\times2\times2$ (right).}
\end{figure}

Since the partial states of the subsystems correspond to Gaussian states, it is possible to compute the different magnitudes by making use of their covariance matrices.
Concretely, we compute the fidelity via the relation given by Paraoanu et al. \cite{Paraoanu1999a}, the entropy via the sympletic eigenvalues \cite{Eisert2010} and the purity via its relation $\purity=1/\sqrt{\text{det}(CM)}$ \cite{Golubeva2014}, where $CM$ represents the covariance matrix.
The results are plotted in Figure \ref{fig:fidelitystate2}.

\section{Correlations in the finite-size model}\label{app:correlations}
In this section, we analyze the behavior of density-density correlations and the characteristic exponents in the finite-size model \eqref{eq:BEmodel}.
We remind here that the results at the thermodynamic limit  are well known, as they correspond to those proven for an ideal Bose gas \cite{Pitaevskii2016}. We, thus, expect to observe similar results when studying the asymptotic behavior.
\subsection{Methods}
Density-density correlations are defined as
\begin{equation}\label{eq:corr}
  \corr(b^\dagger_\textbf{i}\,b_\textbf{i},b^\dagger_\textbf{j}\,b_\textbf{j}) := \expect{b^\dagger_\textbf{i}\,b_\textbf{i}b^\dagger_\textbf{j}\,b_\textbf{j}}-\expect{b^\dagger_\textbf{i}\,b_\textbf{i}} \expect{b^\dagger_\textbf{j}\,b_\textbf{j}}.
\end{equation}
Since the state of interest $\rho_{\{\beta,\,\mu\}}$ is Gaussian, the correlations can be computed via Wick's theorem \cite{Gluza2016, Kraus2009}, given by the expression
\begin{equation}\label{eq:Wicksth}
  \expect{\prod _{k=1}^{m} c_{i_k}}_\beta =\text{Pf}(\Gamma[i_1,\dots, i_{m}]),
\end{equation}
where $c := (b_{\textbf{n}_1},b_{\textbf{n}_1}^\dagger, \dots, b_{\textbf{n}_L},b_{\textbf{n}_L}^\dagger)$ with sites $\textbf{n}_1:=-L_0/2\,\,(1,1,1)$ and \linebreak $\textbf{n}_L:=\textbf{n}_1+L_0-1$; $\text{Pf}(\Gamma[i_1,\dots, i_{m}])=\sqrt{\det(\Gamma[i_1,\dots, i_{m}])}$ is the so-called Pfaffian; and $\Gamma$ has matrix elements
\begin{equation*}
  \big(\Gamma[i_1,\dots,i_{m}] \big)_{a,b} :=
  \begin{cases}
    \phantom{-}\expect{c_{i_a}\ c_{i_b}}_\beta & \text{if}\,a<b, \\
    -\expect{c_{i_b}\,c_{i_a}}_\beta & \text{if}\,a>b, \\
    \phantom{-}0 & \text{otherwise.}
  \end{cases}
\end{equation*}

Applying Equation \eqref{eq:Wicksth} into \eqref{eq:corr}, correlations are given by the covariance matrix elements such that
\begin{equation}\label{eq:fourmoment2}
  \corr(b^\dagger_\textbf{i}\,b_\textbf{i},b^\dagger_\textbf{j}\,b_\textbf{j}) = \expect{b_\textbf{i}^\dagger\,b_\textbf{j}}\,\expect{b_\textbf{i}\,b_\textbf{j}^\dagger}-\expect{b_\textbf{i}^\dagger\,b_\textbf{j}^\dagger}\,\expect{b_\textbf{i}\,b_\textbf{j}} ,
\end{equation}
where the elements in the left-side term correspond to Eqs.~(\ref{eq:covmatfin}-\ref{eq:covmat2}).

\begin{figure}
\centering
\includegraphics[scale=0.5]{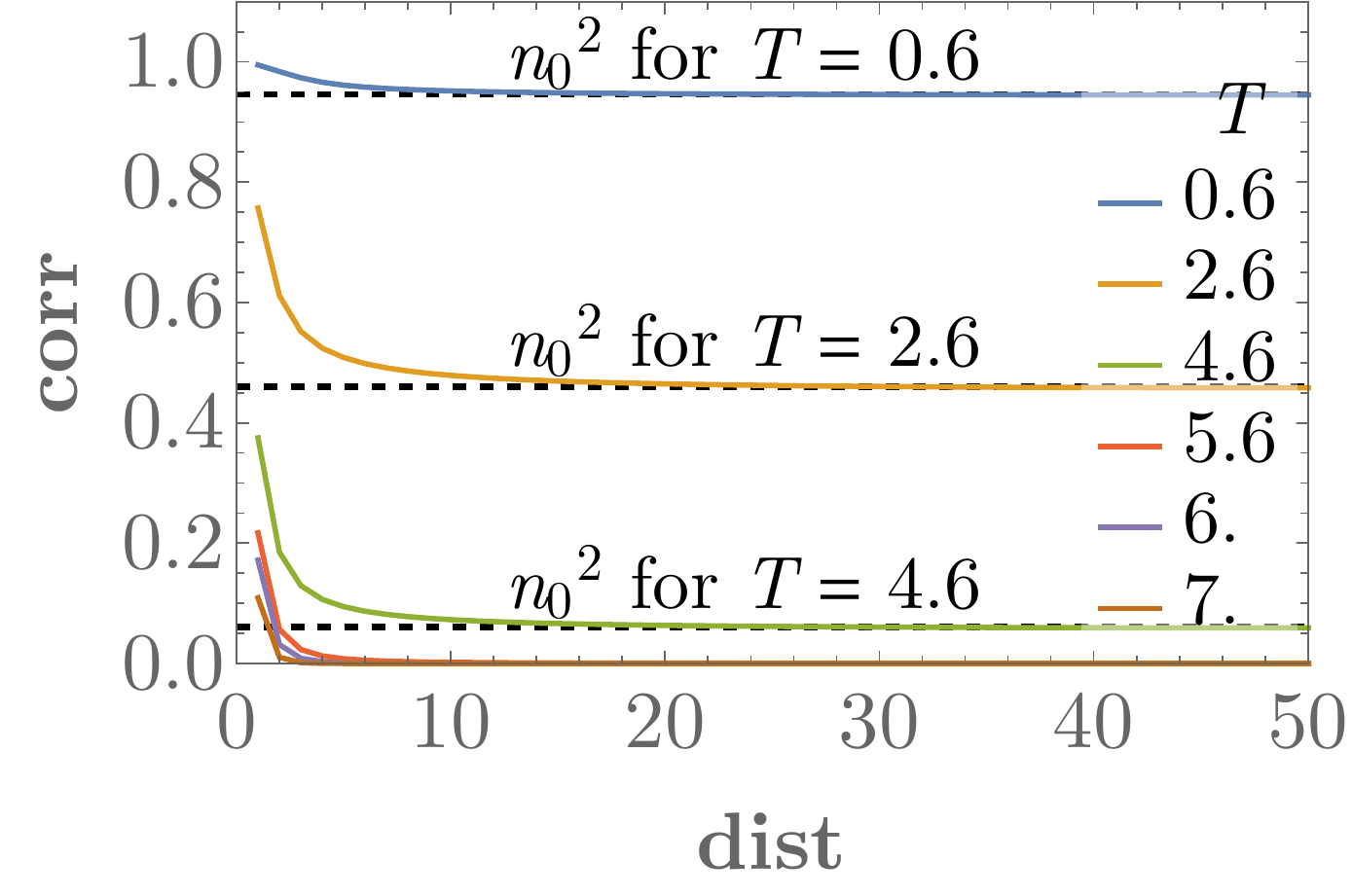}
\caption{\label{fig:correlationsvsdistance}Correlations vs distance for $L=100$ and for temperatures $T={0.6,2.6,4.6,5.6,6.7}$.}
\end{figure}

\subsection{Results}
We focus on the correlations \eqref{eq:fourmoment2} between sites $\textbf{i}:=(i,1,1)$ and $\textbf{j}=(j,1,1)$ as a function of the distance, $\dist := |i-j|$, for a system length $L_0$, a fixed particle density $n=1$ and different temperatures $T$ around the critical temperature $T_c \approx 5.6$.

\begin{figure}[t!]
\centering
\begin{subfigure}[t]{0.48\textwidth}
  \includegraphics[width=0.95\textwidth]{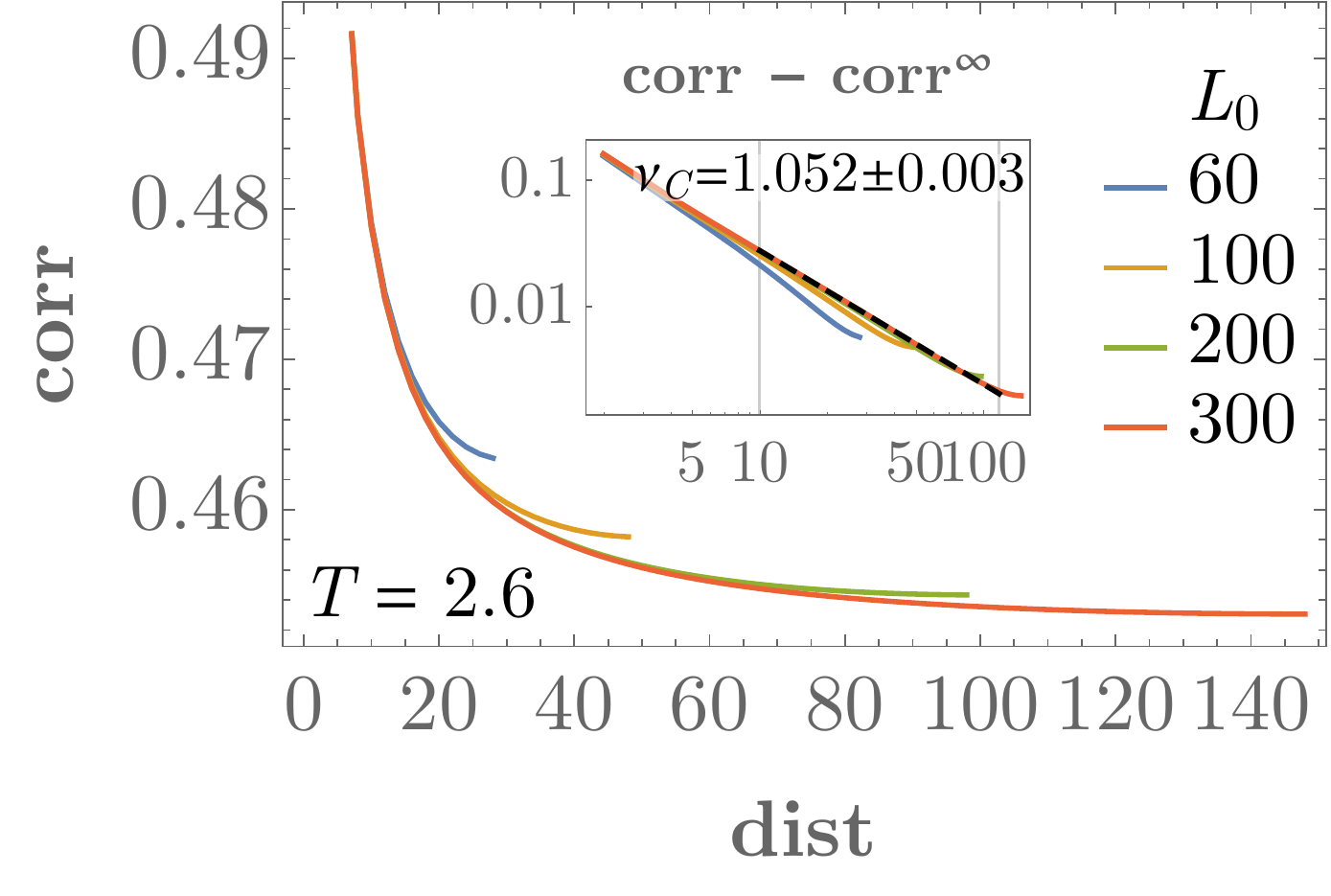}
  \caption{Correlations for $T=2.6<T_c$ with a log-log scaled inset.
  The inset corresponds to the function $\corr - \corr^\infty$, where $\corr^\infty$ is the converging value for $\dist >> 1$.
  The dashed line corresponds to a power-law fit for $L_0=300$ with $\dist\in[10,2\,L_0/5]$.}
\end{subfigure}\hfill
\begin{subfigure}[t]{0.48\textwidth}
  \includegraphics[width=0.92\textwidth]{ExpPL_assumptionfree2.pdf}
  \caption{Exponents for the power-law fits for $T\leq T_c$.}
\end{subfigure}\\
\begin{subfigure}[t]{0.48\textwidth}
  \includegraphics[width=0.97\textwidth]{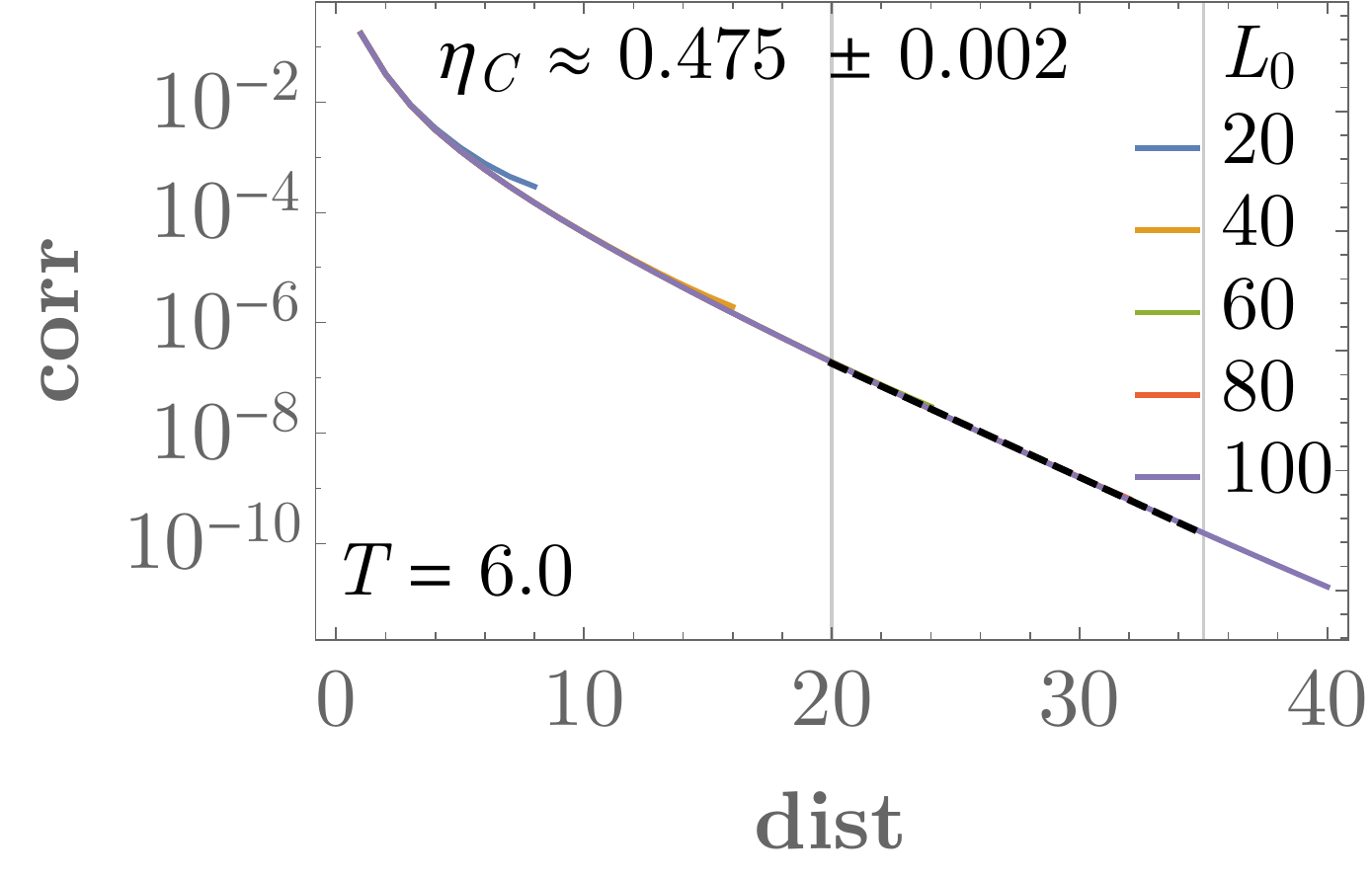}
  \caption{Logarithmic plot of the correlations at $T=6>T_c$. The dashed line corresponds to a exponential fit for $L_0=100$ and data between $\dist=L_0/5$ and $\dist=7\,L_0/20$.}
\end{subfigure}\hfill
\begin{subfigure}[t]{0.48\textwidth}
  \includegraphics[width=0.92\textwidth]{Expexp.pdf}
  \caption{Exponents for the exponential fits for $T>T_c$.
  The dashed line corresponds to a linear fit with exponent $\alpha''$. Notice that the exponent values converge for $L_0\geq 80$.
  Error uncertainty and error bars represent the standard error of the parameters, as obtained from fits.}
\end{subfigure}
\caption{\label{fig:correxponents} Correlation decay above and below the critical temperature.}
\end{figure}

We observe that the correlations behave differently depending on whether we are above or below the critical temperature (see Figure \ref{fig:correlationsvsdistance}), as expected from the known results at the thermodynamic limit \cite{Pitaevskii2016}.
At temperatures $T\leq T_c$, correlations decay to a constant value that asymptotically goes to the square of the ground state density, $n_0^2$, where
\begin{equation*}
  n_0 := (1/L_0^3)\,\expect{b_0^\dagger\,b_0}=\frac{1}{L_0^3}\,\frac{1}{\e^{-\,\mu}-1}.
\end{equation*}
Moreover, we observe that the correlations decay as a power law for $T \in [0.5, T_c]$ (see Figure \ref{fig:correxponents}$.\text{(a)}$), that is,
\begin{equation*}
  \corr(b_{\textbf{i}}^\dagger\,b_{\textbf{i}},b_{\textbf{j}}^\dagger\,b_{\textbf{j}}) - n_0^2 \propto \frac{1}{\dist^{\nu_C}},
\end{equation*}
where the exponent $\nu_C$ is weakly dependent on temperature as $\nu_C \approx 1.05 \pm 0.01$ for $T\leq 4$ and monotonically increases with the temperature for $T\in [4,5.6]$ for $L_0=300$  (see Figure \ref{fig:correxponents}$.\text{(b)}$).
Additionally, the exponent $\nu_C$ converges and its temperature-dependence decreases as the system size increases.
The fits have been obtained by fitting the correlations to a polynomial function with a free constant: $a/\dist^b+c$ with $\dist\in[10,2\,L_0/5]$.

Regarding the results for $T \leq 0.5$, the exponents were too noisy for the system sizes considered and, thus, they are unreliable and not included.
This is in great part due to the three-dimensional nature of the system and the slow convergence at low temperatures, which challenges to obtain reliable results for the sizes we were able to study.\\

For temperature $T>T_c$, we observe that correlations decay exponentially to zero (see Figure \ref{fig:correxponents}$.\text{(c)}$), that is,
\begin{equation*}
  \corr(b_{\textbf{i}}^\dagger\,b_{\textbf{i}},b_{\textbf{j}}^\dagger\,b_{\textbf{j}}) \propto \e^{-\,\eta_C\dist},
\end{equation*}
where the exponent $\eta_C$ increases with the temperature $T$.
In particular, it increases linearly, i.e., $\eta_C=\alpha'' T+\beta''$ with a factor $\alpha''\approx 0.759\pm 0.002$ and with values $\eta_C\in[0,1.2]$ for $T\in[5.6,7]$ (see Figure \ref{fig:correxponents}$.\text{(d)}$).
The fits have been computed for data with $\dist\in[L_0/5,7\,L_0/20]$.\\

We highlight that the results for large enough system sizes are similar to the theoretical results proven for an ideal Bose gas at the thermodynamic limit. In fact, in our model we recover the continuum in the thermodynamic limit if we replace $\cos k$ by $1-k^2/2$, that is, if only the modes with small $k$ are occupied, which certainly occurs below $T_c$.
For instance, it is known that density-density correlations decay to $n_0^2$ with exponent $\nu_C^{\infty}=1$ for $T\leq T_c$ at the thermodynamic limit.
Thus, we can clearly see that our numerical results for low temperature are converging to the theoretical ones, as the exponent $\nu_C \approx 1.05 \pm 0.01$ for low temperatures and temperature-dependence decreases with system size.
Moreover, it is known that correlations decay exponentially with an exponent that increases with temperature for $T>T_c$, behavior that also coincides with our numerical results.
The theoretical results can be obtained by combining the results for one-body correlations (page 25, section 3.2 of Pitaevskii and Stringary \cite{Pitaevskii2016}) with the relation between two-body and one-body correlations (see Eq.~\eqref{eq:fourmoment2}, or section 3.3 of Pitaevskii and Stringary~\cite{Pitaevskii2016}).

\end{document}